\documentclass[aps,prl,10pt,twocolumn,a4paper,nofootinbib,preprintnumbers,superscriptaddress]{revtex4-2} 

\usepackage{bm}
\usepackage{amsmath,amssymb}
\usepackage{amsfonts}
\usepackage{layouts}
\usepackage{mathtools}
\usepackage{dsfont}
\usepackage{graphicx}
\usepackage{float}
\usepackage{subfigure} 
\usepackage{verbatim}
\usepackage{amsfonts}
\usepackage{bbold}
\usepackage{dcolumn}
\usepackage{bm}
\usepackage{color}
\usepackage[dvipsnames]{xcolor}
\usepackage{listings}
\definecolor{blueprl}{RGB}{46,48,146}

\usepackage{mathrsfs}
\usepackage{filecontents}
\usepackage{times} 
\newcommand{\bra}[1]{\mbox{$\langle #1 |$}}
\newcommand{\ket}[1]{\mbox{$| #1 \rangle$}}

\newcommand{\xdownarrow}[1]{%
  {\left\downarrow\vbox to #1{}\right.\kern-\nulldelimiterspace}
}

\makeatletter
\newcommand*{\balancecolsandclearpage}{
  \close@column@grid
  \clearpage
  \twocolumngrid
}
\makeatother

\usepackage{gensymb}
\usepackage[breaklinks=true]{hyperref}
\hypersetup{
colorlinks   = true, 
urlcolor     = blue, 
linkcolor    = blue, 
citecolor    = blue 
}
\usepackage{comment}
\usepackage[normalem]{ulem}
\usepackage{xcolor}
\usepackage{graphicx}
\usepackage{cleveref}

\usepackage{cleveref}
\crefname{equation}{Eq.}{Eqs.}
\Crefname{equation}{Equation}{Equations}
\crefname{figure}{Fig.}{Figs.}
\Crefname{figure}{Figure}{Figures}
\crefname{figure}{Fig.}{Figs.}
\Crefname{figure}{Figure}{Figures}
\crefname{section}{Sec.}{Secs.}
\Crefname{section}{Section}{Sections}
\crefname{appendix}{Appendix}{Appendices}
\Crefname{appendix}{Appendix}{Appendices}
\crefname{table}{Table}{Tables}
\Crefname{table}{Table}{Tables}

\usepackage{enumitem,amssymb}
\newlist{todolist}{itemize}{2}
\setlist[todolist]{label=$\square$}
\usepackage{pifont}

\usepackage[T1]{fontenc}
\usepackage{lmodern} 

\usepackage[most]{tcolorbox}

\tcbset{textmarker/.style={%
        enhanced,
        parbox=false,boxrule=0mm,boxsep=0mm,arc=0mm,
        outer arc=0mm,left=3mm,right=3mm,top=7pt,bottom=7pt,
        toptitle=1mm,bottomtitle=1mm,oversize}}

\newtcolorbox{hintBox}{textmarker,
    borderline west={6pt}{0pt}{yellow},
    colback=yellow!10!white}

\newtcolorbox{noteBox}{textmarker,
borderline west={-0.5pt}{-0.5pt}{white},
colback=blue!10!white}
\newtcolorbox{importantBox}{textmarker,
borderline west={-1pt}{-1pt}{white},
colback=red!10!white}

\begin{document}

\title{All-photonic quantum key distribution beyond the single-repeater bound}

\author{Matthew S. Winnel}\email{mattwinnel@gmail.com}\affiliation{Toshiba Europe Ltd, 208 Cambridge Science Park, Cambridge CB4 0GZ, United Kingdom}
\author{Sergio Ju\'arez}\affiliation{Toshiba Europe Ltd, 208 Cambridge Science Park, Cambridge CB4 0GZ, United Kingdom}
\affiliation{Escuela de Ingeniería de Telecomunicación, Department of Signal Theory and Communications, University of Vigo, Vigo E-36310, Spain}
\author{Chithrabhanu Perumangatt}\affiliation{Toshiba Europe Ltd, 208 Cambridge Science Park, Cambridge CB4 0GZ, United Kingdom}
\author{Taofiq Paraiso}\affiliation{Toshiba Europe Ltd, 208 Cambridge Science Park, Cambridge CB4 0GZ, United Kingdom}
\author{R. Mark Stevenson}\affiliation{Toshiba Europe Ltd, 208 Cambridge Science Park, Cambridge CB4 0GZ, United Kingdom}

\date{\today}

\begin{abstract}
Quantum protocols require classical signaling, and when classical signals propagate faster than quantum ones, standard rate–loss limits can be surpassed. We introduce an all-photonic measurement-device-independent quantum key distribution protocol that exceeds the single-repeater bound without error correction. When quantum signals travel at two-thirds the classical speed, the key rate scaling approaches $\eta^{2/5}$. We propose a single-rail, temporally multiplexed architecture that extends twin-field–type protocols to multiple nodes and surpasses their key rate without ideal quantum memories.

\end{abstract}

\maketitle

\textit{Introduction} — Quantum key distribution (QKD)~\cite{Bennett_2014,Scarani_2009,Pirandola_2020,Diamanti_2016} enables information-theoretically secure communication by exploiting the principles of quantum mechanics. Long-distance implementations, however, remain fundamentally constrained by the exponential decay of transmission rates induced by channel loss~\cite{Takeoka_2014,Pirandola_2017}. Among purely optical, memoryless approaches, twin-field (TF) QKD~\cite{Lucamarini_2018} achieves the best-known rate–loss scaling by exploiting single-photon interference at an untrusted relay, yet offers no simple path to surpass its scaling without introducing quantum memories or error-corrected repeaters. The conventional route to overcoming these limitations relies on quantum repeater architectures~\cite{Muralidharan_2016}, which typically require long-lived quantum memories or fault-tolerant quantum operations—capabilities that remain technologically demanding and far from widespread deployment.

Quantum teleportation~\cite{PhysRevLett.70.1895} transfers a quantum state using shared entanglement and classical communication, but its practical implementation normally requires waiting for the classical message that specifies the corrective operation. Such timing constraints are a central limitation of teleportation-based protocols, including multi-node QKD and repeater architectures.

Our key observation is that classical information may propagate significantly faster than quantum signals. By exploiting this speed advantage, the receiver can obtain the classical correction information before the quantum signal arrives, removing the need for long-lived memories and eliminating the usual feedforward delay. We apply this mechanism here to demonstrate improved key-rate scaling in untrusted quantum networks—surpassing the performance of TF-QKD and even exceeding the single-repeater bound without requiring perfect quantum memories or error correction.

\begin{figure*}
\centering
\includegraphics[width=0.9\textwidth]{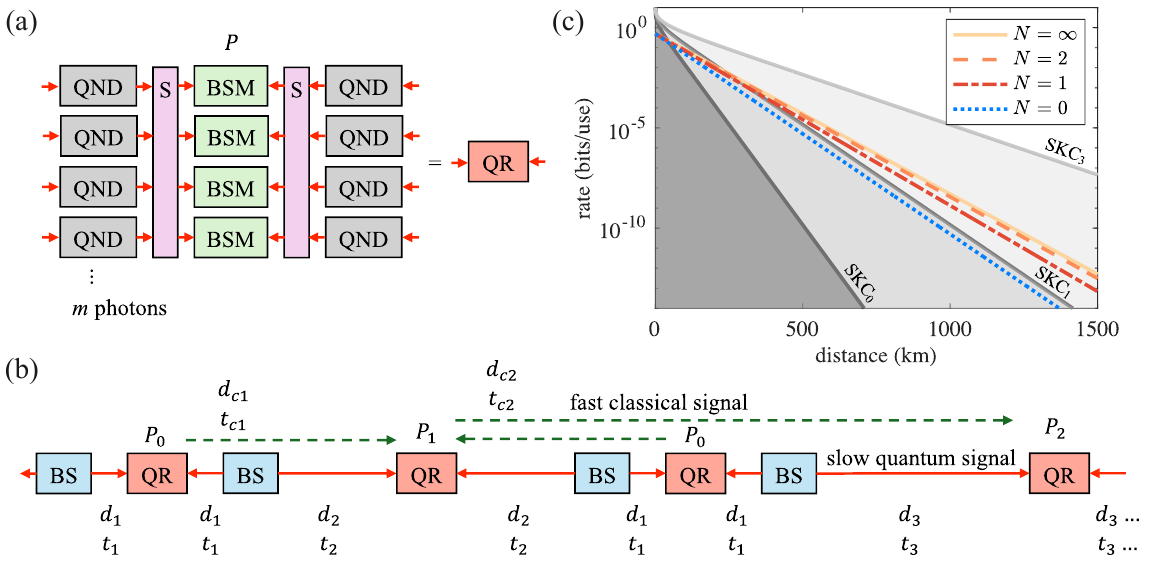} \caption{(a) Repeater protocol from Ref.~\cite{Azuma_2015}, where $ m $ dual-rail qubits (photons) undergo quantum non-demolition (QND) measurements at a quantum repeater (QR) node. A fast optical switch (S) routes photons to a Bell state measurement (BSM), with total success probability $ P $ determining the rate. (b) Schematic of the ideal protocol using fast classical communication. Photons are released simultaneously at the Bell states (BS), with quantum link distances $ d_i $, with associated times $ t_i $, and classical counterparts $ d_{ci} $, $ t_{ci} $. Classical signals ``catch up'' before the quantum signal reaches the next repeater, e.g., $ d_1 + d_2 = d_{c1} $, $ t_2 = t_1 + t_{c1} $. (c) Key rates (bits/use) vs. total distance, assuming channel loss rate $ \alpha = 0.2 $ dB/km and incorporating our fast-classical-communication technique. Here, $N$ indicates the number of nested repeater levels, with $N=0$ corresponding to the original memoryless protocol in Ref.~\cite{Azuma2015}. The quantum speed is $ 2/3 $ of the classical speed $ c $, with rates approaching $ \eta^{2/5} $ (yellow solid). As $N\to\infty$, $\text{SKC}_0$ is overcome at 41.2 km and $\text{SKC}_1$ at 230.6 km.}\label{fig:protocol_ideal}
\end{figure*}

Ref.~\cite{Azuma2015} introduced a memoryless measurement-device-independent (MDI) QKD protocol based on dual-rail qubits~\cite{eczoo_dual_rail, Winnel_2022} and a single untrusted relay node, shown in \cref{fig:protocol_ideal}(a). Alice and Bob each send photons to the relay, where a quantum non-demolition (QND) measurement registers their arrival with probability  $\sqrt{\eta},$ with $\eta$ denoting the end-to-end transmissivity. Detected photons are then paired for a Bell-state measurement. This node-receives-photon (NRP) architecture~\cite{https://doi.org/10.1002/qute.201900141} eliminates the need for quantum memories, distinguishing it from node-sends-photon (NSP) approaches. With sufficient spatial multiplexing, the protocol achieves a key rate $K_0 = \frac{\sqrt{\eta}}{2}$ bits per channel use. Whether a scheme of this type could surpass the single-repeater bound, $\mathrm{SKC}_1 = -\log_2(1 - \sqrt{\eta})~\cite{Pirandola_2019},$ has remained unresolved. In this work, we show that it can, once the speed difference between classical and quantum signals is incorporated into the architecture.

We introduce an all-photonic MDI-QKD protocol that surpasses the single-repeater bound by exploiting the speed gap between quantum and classical signals, together with optimally spaced intermediate nodes. Quantum signals may propagate at $c_{\mathrm q}$ while classical messages travel at $c_{\mathrm c}$ such that $c_{\mathrm q}<c_{\mathrm c}$ then classical information can overtake or ``catch up to'' the slower quantum pulses (for instance, classical signals may propagate through free space or low-latency media such as hollow-core fibers, and quantum signals in silica fibers~\cite{10526513,Petrovich2025}). We derive the secret-key rate for an ($N$)-nested pure-loss channel and show that, under these conditions, the rate approaches $K_\infty = \tfrac{1}{2} \eta^{\frac{c_{\mathrm q}}{c_{\mathrm c}+c_{\mathrm q}}}$ bits per use, exceeding the single-repeater bound without requiring perfect quantum memories or error correction. In the extreme limit $c_{\mathrm c} \gg c_{\mathrm q}$, i.e., when quantum signals propagate much more slowly than classical ones, the asymptotic rate saturates at $K_\infty = \tfrac{1}{2},$ indicating that arbitrarily long-distance communication is, in principle, achievable at a constant key rate, provided that the quantum signals can be slowed without increasing their attenuation coefficient (in dB/km). Straighter classical paths (such as line-of-sight free-space channels) can further reduce classical travel time and enhance this effect.

\textit{Fast classical communication} — The single-node, memoryless QKD protocol of Azuma et al.~\cite{Azuma2015} serves as the building block for our multi-node scheme, shown in \cref{fig:protocol_ideal}(b). As illustrated in \cref{fig:protocol_ideal}(c), the ideal key rate of Azuma’s protocol approaches $K = \frac{1}{2}\sqrt{\eta}$ (N = 0, no nesting, squares in blue). This exceeds the repeaterless bound $\text{SKC}_0$ and follows a similar scaling to the single-repeater bound $\text{SKC}_1$, although it does not surpass it~\cite{Pirandola_2019}.

To overcome this limitation, we extend the protocol to a multi-node network, also shown in \cref{fig:protocol_ideal}(b), by exploiting the speed difference between classical ($c_\text{c}$) and quantum ($c_\text{q}$) signals. All photons are emitted from the Bell states (BS) simultaneously. By optimally positioning relay nodes, classical messages can be timed to arrive concurrently with the quantum signals. 

We compute the secret key rate for nested architectures: with one level of nesting ($N = 1$), the rate is $K_1 = \frac{1}{2}\eta^{c_\text{c}/(3c_\text{c} - c_\text{q})}$; in the limit of many nesting levels, the rate approaches $K_\infty = \frac{1}{2}\eta^{c_\text{q}/(c_\text{c} + c_\text{q})}$. Additional rate expressions are derived in the Supplemental Material. In \cref{fig:protocol_ideal}(c), we plot these results assuming $c_\text{q} = \frac{2}{3}c_\text{c}$, with $c_\text{c} = c$, the speed of light in vacuum.

\textit{Our proposed implementation} — 
While the dual-rail architecture of Ref.~\cite{Azuma2015} relies on QND measurements, single-rail protocols based on single-photon interference, such as TF-QKD~\cite{Lucamarini_2018}, avoid this requirement and are well suited for scalable, all-photonic implementations. Building on this foundation, we describe the architecture that realizes our fast-classical-communication advantage. Our construction is a multi-node extension of TF-QKD using temporal multiplexing, in which each node employs only linear-optical operations and classical feedforward. The temporal multiplexing is synchronized using lossy optical buffers, effectively very poor memories, which nevertheless suffice for coordinating the required timing. This design surpasses single-repeater scaling through fast classical signaling without ideal quantum memories, and its performance would improve further as higher-quality memories become available. While the architecture supports further nesting, we focus on a single level to illustrate the core construction.

In contrast to previous multi-node extensions of TF-QKD~\cite{PhysRevA.102.042614, Rozp_dek_2019}, our scheme provides several key advantages. It is fully optical, avoiding depolarization, weak nonlinearities, and hybrid qubit encodings; it supports GHz-level repetition rates; and it remains robust even when implemented with lossy quantum memories such as fiber loops or free-space optical buffers. 

\begin{figure*}
\centering
\includegraphics[width=0.95\textwidth]{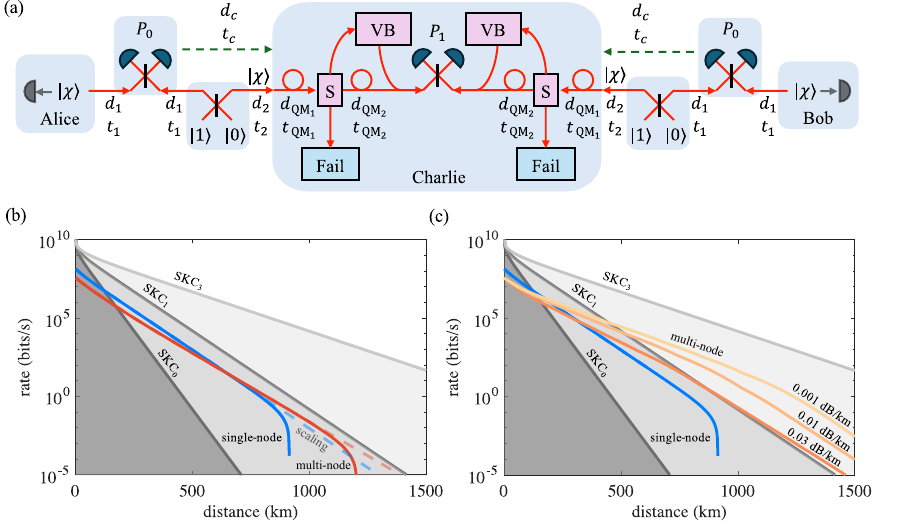} \caption{(a) Multi-node quantum repeater protocol extending single-photon interference schemes beyond single-repeater scaling. In the entanglement-based version, TMSV states are distributed, whereas in the prepare-and-measure version, Alice and Bob measure this ``fictitious'' entanglement immediately. Single-photon states serve as an additional resource for efficient long-distance teleportation. Photons are transmitted over quantum channels (distances $d_1, d_2$, with associated times $t_1, t_2$) with total distance $L = 2(2d_1 + d_2)$. Fixed optical buffers (red loops) act as lossy quantum memories (distances $d_{\text{QM}_1}$, times $t_{\text{QM}_1}$), while optical switches (S) route signals based on measurement outcomes, with fixed buffer delay $t_{\text{QM}_2}$ and variable buffers (VB) for synchronisation. Fast classical communication (distances $d_c$, times $t_c$) enhances key rates despite memory losses.  (b) Key rates vs. total physical distance for quantum memory loss rate 0.2 dB/km, compared with a single-node scheme and fundamental bounds~\cite{Pirandola_2017,Pirandola_2019}. The rate of the single-node protocol is overcome at 652.4 km by the rate of the multi-node protocol. $\text{SKC}_1$ would be overcome at 1621 km if not for dark counts. (c) Same as (b), but assuming quantum-memory loss rates below the break-even threshold (i.e., $< 0.05$ dB/km). Quantum error correction (QEC) inside $d_{\text{QM}_1}$ could bring loss rates into this regime. Parameters: 1 GHz repetition rate, 0.2 dB/km channel loss, TMSV parameter $ \chi = 0.25 $, 99$\%$ switching efficiency, 93$\%$ detector efficiency, and 0.01 Hz dark counts~\cite{Marsili_2013}. Dark counts limit maximum distance (dashed line shows performance without dark counts).  }\label{fig:protocol}
\end{figure*}

We now outline how the scheme can be constructed in practice, as shown in~\cref{fig:protocol}(a). For the purpose of analysis, we imagine that Alice and Bob distribute two-mode squeezed vacuum (TMSV) states through the network (i.e., single-rail entanglement, e.g. spontaneous parametric down conversion) and evaluate performance using the product of the system's rate and its reverse coherent information (RCI)~\cite{Garc_a_Patr_n_2009}, which provides a lower bound on an achievable secret-key rate. The TMSV state is given by $\ket{\chi} = \sqrt{1 - \chi^2} \sum_{n=0}^{\infty} \chi^n \ket{n}\ket{n}$, where $ \chi = \tanh r $ is the squeezing parameter, $ r $ is the squeezing amplitude, $ \bar{n} = \sinh^2 r $ is the mean photon number, and $ \nu = \cosh 2r $ is the quadrature variance. The states $ \ket{n} $ denote Fock number states. 

In practice, Alice and Bob need not store this entanglement in long-lived quantum memories. They may do so in principle, but if such memories are unavailable, they can simply measure their halves of the TMSV states immediately. This converts the scheme into a prepare-and-measure (PM) protocol whose classical statistics are exactly those obtained by measuring the fictitious entanglement. The entanglement-based description is therefore used only as a standard theoretical framework for lower-bounding the secret-key rate; the implemented protocol can be entirely measurement-based and memory-free. A successful repeater chain then corresponds to the generation of strong correlations between Alice and Bob, enabling secure key extraction.

The channel and buffer transmission efficiencies are
$\eta_i = 10^{-\alpha d_i/10}$ and $\eta_{\text{QM}_i} = 10^{-\alpha_\text{QM} d_{\text{QM}_i}/10}$, where $\alpha$ and $\alpha_\text{QM}$ are the loss  coefficients (in dB/km) of the quantum channel and optical memory, respectively, and $\eta_\text{switch}$ is the transmissivity of a single optical switch. The pulse duration is $\tau = 1/R_\text{rep}$, and the variable delay buffer is implemented digitally with delays $b^k \tau$ for integers $k = 0, 1, 2, \ldots$, where $b$ is the base of the delay register. We assume every delay in the digital buffer has a lossy switch increasing the total loss.

One mode of each TMSV state is retained locally, while the other mode is sent through a lossy quantum channel of length $ d_1 $. Simultaneously, two copies of single-photon entangled states, created by interfering a single photon with vacuum on a 50:50 beamsplitter, are generated at intermediate relay nodes, as illustrated in~\cref{fig:protocol}(a).

One half of each entangled state is sent to the nearest intermediate node over distance $d_1$, while the other travels distance $d_2$ before entering lossy quantum memories of lengths $d_{\text{QM}_1}$ and $d_{\text{QM}_2}$. These memories enable entanglement swapping by synchronizing arrival times.

The intermediate node performs a Bell-state measurement (BSM) with success probability $P_0 \approx \eta_1 / 2$ in the small-$\chi$ limit, teleporting Alice and Bob's modes into quantum memories, with additional loss from the $d_2$ segments. This yields near-optimal rate-distance scaling $\propto \eta_1$~\cite{winnel2021overcomingrepeaterlessboundcontinuousvariable}. Effectively, Alice's mode is teleported to the start of channel $d_2$ (marked ``$\ket{\chi}$'' in~\cref{fig:protocol}(a)), and likewise for Bob. High teleportation fidelity is maintained for small $\chi$. The value of $d_2$ can be optimized for key rate and may be zero with ideal quantum memories.

The teleported states then propagate through lossy quantum memories and are synchronized before detection at Charlie's central station. Optical switches (with each traversal of a loop in the digital buffer incurring one switch use) route each successfully teleported state into fixed or variable buffers depending on whether a matching state from the opposite side is already stored, discarding failed rounds in real time. Once both sides have heralded success, a variable buffer synchronizes the two states so that they arrive simultaneously at Charlie's BSM. To balance losses at Charlie's beamsplitter, a variable beamsplitter is used on the less lossy side.

Upon a successful repeater chain, Alice and Bob share a single-rail Bell state
\begin{equation}
\ket{\psi}_\text{final} \approx \frac{1}{\sqrt{2}}\bigl(\ket{10} + e^{i\phi}\ket{01}\bigr).
\end{equation}
A secret key may then be extracted using a protocol analogous to TF-QKD. In practice, the protocol can be implemented in a PM form using phase-randomized weak coherent states. Standard decoy-state techniques~\cite{Lucamarini_2018} can be employed to estimate the single-photon contributions, enabling a security analysis following that of conventional TF-QKD.

\textit{Secret key rate} — We first derive ideal achievable rates for pure-loss channels in the small-$\chi$ limit, then present numerical results including finite $\chi$, detector inefficiencies, and dark counts.

The secret key rate (SKR) is
\begin{align} 
\text{SKR} &= \frac{r R}{\tau}, 
\end{align}
where $r$ is the raw key rate (an achievable rate is lower-bounded by the RCI), and $R$ is the success probability of the full repeater protocol.

The repeater rate is
\begin{align}
    R &= \frac{1}{\overline{Z}_0^{(m)}(P_0)} \frac{1}{\overline{Z}_1 \left( P_1 \right)},
\end{align}
where $P_0$ and $P_1$ are the success probabilities of the initial and final BSMs, and $\overline{Z}_0^{(m)}$, $\overline{Z}_1$ are the corresponding average wait times. The fixed buffer duration $t_{\text{QM}_2} = m\tau$ limits maximum storage time, but the rate includes this constraint (see~\cite{PhysRevLett.98.060502,PhysRevA.100.032322}).

Photons are stored for a total duration $t_{\text{QM}_1}+t_{\text{QM}_2}$, where $t_2+t_{\text{QM}_1}=t_1+t_c$ compensates the classical--quantum propagation mismatch so that the buffered quantum state is released upon arrival of the classical feedforward, and $t_{\text{QM}_2}$ provides synchronization with the neighboring quantum signal and the classical message from the opposite node.

We assume the same loss rate (dB/km) for all buffers: $d_{\text{QM}_1}$, $d_{\text{QM}_2}$, and the variable buffer (VB). Excluding switch losses, VB loss is no greater than that of $d_{\text{QM}_2}$, since its storage time is shorter. However, the VB adds extra loss from optical switching.

In the small-$\chi$ limit, the ideal secret key rate is (see the Supplemental Material): 
\begin{align}
\lim_{\chi \to 0} \text{SKR} &= \chi^2 \, \eta_1 \, \eta_2 \, \eta_{\text{QM}_1} \, \eta_{\text{QM}_2} \, \eta_\text{switch}^{b \lceil \log_b(m) \rceil + 1}  \nonumber
\\ &\;\;\;\; \times \frac{ \left( \frac{\eta_1}{2} + 2 \left( 1 - \frac{\eta_1}{2} \right)^{m + 1} - 2 \right)}{\tau \left( \eta_1 + 2 \left( 1 - \frac{\eta_1}{2} \right)^{m + 1} - 3 \right)} \text{ bits/s}.
\end{align}

In~\cref{fig:protocol}(b), we plot the numerically obtained secret-key rates as a function of distance, incorporating finite $\chi$ as well as detector inefficiencies and dark counts, assuming optical buffers with fast switching serve as quantum memories with loss rate $\alpha_\text{QM} =0.2$. In~\cref{fig:protocol}(c), we show the same setting as in (b), but with quantum-memory loss rates below the break-even point. We set $\tau = 10^{-9}$ s, channel loss $\alpha = 0.2$ dB/km, and propagation speeds $c_{\text{q}} = \frac{2}{3}c$, $c_{\text{c}} = 0.9997c$, and $c_{\text{QM}} = c_{\text{q}}$ for $\alpha_\text{QM} = 0.2$, otherwise $c_{\text{QM}} = c$. Parameters $m$, $\chi$, and node positions are optimized for each case.

 This architecture enables secure communication across a chain of nodes and allows the single-repeater bound to be exceeded using only fast classical signaling, nevertheless, it does not yet achieve the optimal scaling of full quantum-repeater networks~\cite{Pirandola_2019}. With ideal memories (e.g., via error correction), the multi-node protocol approaches the optimal $\eta^{1/4}$ scaling. Approaching those ultimate limits will require low-loss quantum memories. Once available, the scheme can be extended while remaining fully photonic. In particular, the memories may be realized using one-way repeater modules~\cite{Azuma_2015, Lee_2019, munro2013quantumcommunicationnecessityquantum}. In such architectures, the necessary repetition and error-correction operations are performed locally within each repeater node, effectively functioning as a quantum memory. This removes the otherwise prohibitive requirement of executing error correction every $\sim 15$ km along the transmission channel, rendering long-distance deployment far more practical.

The choice between NRP and NSP configurations depends on whether quantum memories are available and, if so, on the level of memory loss, as characterized by their attenuation coefficient. When memories are not available or if memory loss is high, the NRP configuration combined with fast classical communication is advantageous, whereas decreasing memory loss shifts the balance toward the NSP configuration. Assuming a channel attenuation of 0.2 dB/km, the break-even point occurs at a memory attenuation of roughly 0.1 dB/km or lower. The precise break-even thresholds depend on the relative signal speeds and are derived in the Supplemental Material. Encouragingly, attenuation below 0.1 dB/km has been experimentally demonstrated in broadband hollow-core fibers~\cite{Petrovich2025}, with the same work projecting longer-term reductions to 0.033 dB/km and potentially 0.018 dB/km. Importantly, our all-optical scheme not only surpasses the rate–loss scaling of MDI-QKD and TF-QKD without quantum memories, but also extends the maximum achievable communication distance by effectively functioning as a relay~\cite{Jacobs_2002,Zou_2025}.

\textit{Conclusion} — We have shown that the speed difference between quantum and classical communication can be harnessed to improve quantum teleportation and entanglement swapping. By allowing classical information to configure operations before or when quantum states arrive, this timing asymmetry reduces reliance on active quantum storage. This principle offers a new approach to designing scalable, memory-light quantum communication protocols—particularly in photonic networks where propagation delays dominate.

To demonstrate its usefulness, we proposed an all-photonic repeater architecture that surpasses single-repeater scaling without quantum memories. We first introduced a general protocol based on dual-rail qubits and QND measurements. To improve practicality, we then developed a refined version using single-photon interference and single-rail qubits, eliminating the need for QNDs. This implementation integrates single-photon sources, lossy quantum memories, fast optical switches, and Bell-state measurements. It outperforms single-node schemes such as TF-QKD in both key rate and range, while remaining compatible with protocols like TF-QKD and mode-pairing QKD~\cite{Zeng2022}. Its fully optical design avoids light–matter interactions (minimizing dephasing and depolarization) and sidesteps weak nonlinearities, enhancing practicality. Phase mismatch, however, remains a key challenge for maintaining interference fidelity.

Our results underscore the importance of low-loss quantum memories ($< 0.05$ dB/km) for achieving full quantum repeater scaling, as fast classical communication alone has limits. Nonetheless, our approach enables multi-node protocols to outperform single-node ones even without ideal memories or error correction. We hope this work inspires new architectures that exploit speed disparities, with future directions including light slowing for more efficient repeaters. 

\textit{Acknowledgements}  — We acknowledge support from the European Union's Horizon Europe Framework Programme under the Marie Sk\l{}odowska Curie Grant No.101072637, Project Quantum-Safe Internet (QSI).

%

\makeatletter
\renewcommand \thesection{S\@arabic\c@section}
\renewcommand \thetable{S\@arabic\c@table}
\renewcommand \thefigure{S\@arabic\c@figure}
\makeatother
\renewcommand{\theequation}{S.\arabic{equation}}

\clearpage
\onecolumngrid

\setcounter{figure}{0}
\setcounter{equation}{0}
\setcounter{table}{0}
\setcounter{section}{0}
\setcounter{page}{1}

\begin{center}
{\LARGE\textbf{Supplemental Material}}\\[0.3em]
{\LARGE All-photonic quantum key distribution beyond the single-repeater bound}\\[0.2em]
\end{center}

\vspace{0.5em}

\begin{center}
Matthew S. Winnel, Sergio Ju\'arez$^{1,2}$, Chithrabhanu Perumangatt, Taofiq Paraiso, and R. Mark Stevenson\\[0.3em]
$^1$\textit{Toshiba Europe Ltd, 208 Cambridge Science Park, Cambridge CB4 0GZ, United Kingdom}\\
$^2$\textit{Escuela de Ingeniería de Telecomunicación, Department of Signal Theory and Communications, University of Vigo, Vigo E-36310, Spain}
\end{center}

\onecolumngrid

\section{Derivation of ideal key rate scaling}

In this section, we derive the ideal secret key rates of the dual-rail protocol using faster-than-fibre classical communication, as shown in Fig.~1 of the main text.

\subsection{$N=0$}

We begin with the simplest protocol involving a single quantum repeater, as depicted in the following. We employ the quantum repeater design proposed by Munro et al., which could readily be generalised to include the single-shot quantum error detection protocol~\cite{winnel2021overcomingrepeaterlessboundcontinuousvariable}:
\begin{figure}[H]
\centering
 \includegraphics[width=0.7\linewidth]{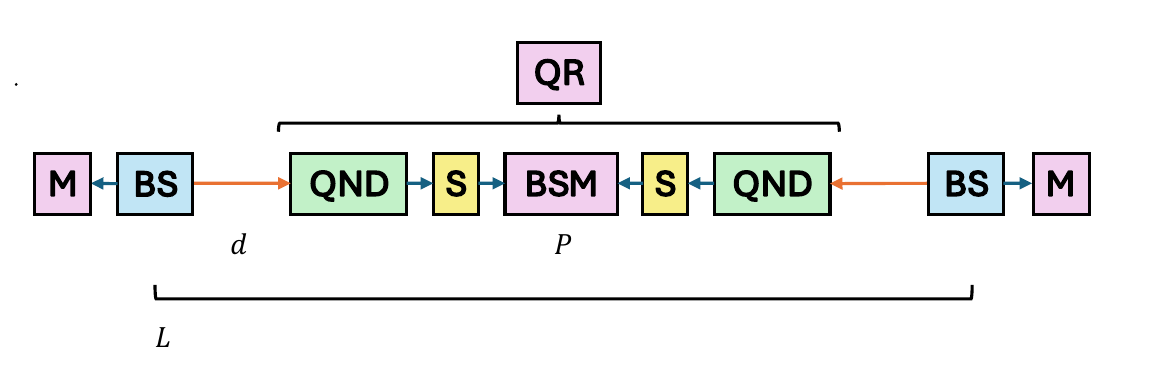}\label{fig:N1}
\end{figure}
\noindent
where we define measurement (M), Bell state (BS), quantum non-demolition measurement (QND), optical switches (S), Bell-state measurement (BSM), quantum repeater (QR), total distance in km ($L$), distance of the link in km ($d$), and probability of the BSM ($P$). The red arrows signify lossy channels with transmissivity $\eta = 10^{-\alpha d/10}$ where $\alpha$ is the attenuation coefficient, indicating signal loss in dB/km.

Assuming a large degree of multiplexing, $m \to \infty$, the rate in bits per use is given by the probability of success divided by two (since dual-rail qubits use the channel twice), that is, $K = P/2 = \eta_d/2 = 10^{-\alpha d/10}/2 = \sqrt{\eta}$/2, where $d = L/2$, with $L$ representing the total distance and $\eta$ denoting the overall transmissivity of the channel length $L$.

Hereafter, to simplify the illustrations we refer to the entire repeater node simply as QR.

\subsection{$N=1$}

The next simplest protocol with one level of nesting of quantum repeaters is shown in the following diagram:
\begin{figure}[H]
\centering
 \includegraphics[width=0.7\linewidth]{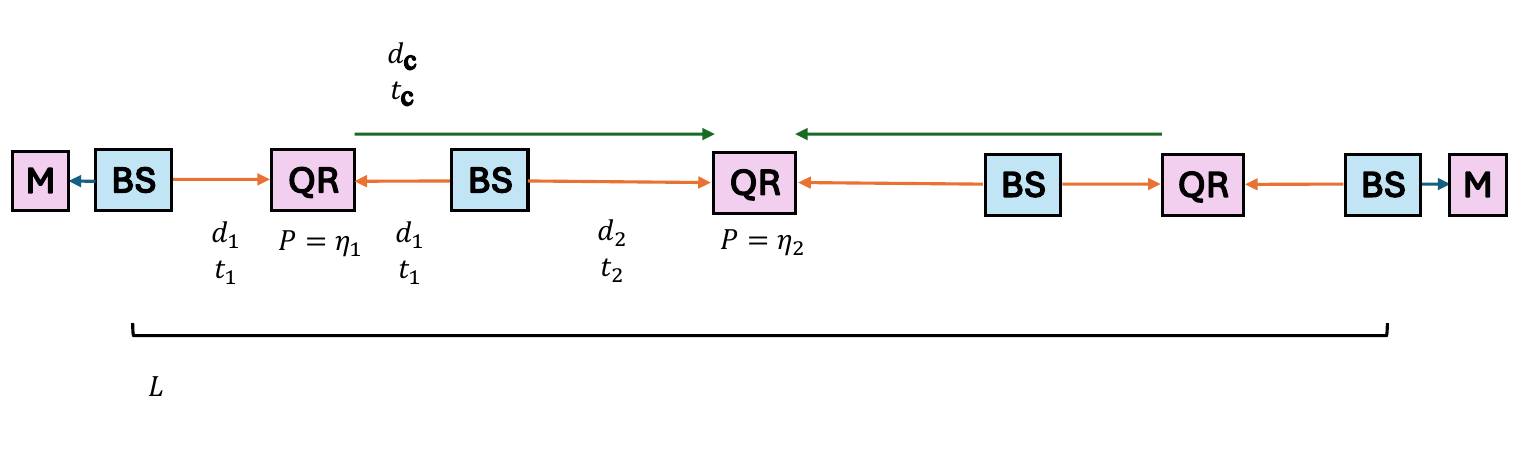}\label{fig:N2}
\end{figure}

All Bell states (BS) emit their photons simultaneously. Alice and Bob may either measure the entanglement immediately or prepare states from an ensemble equivalent to having performed such a measurement.

We require that
\begin{align}
    d_\text{c} &= d_1 + d_2,
\end{align}
where $d_\text{c}$ is the distance classical communication travels which must be equal to the distance of left plus right travelling quantum signals, as shown.

We choose the position of the relays so that the classical signal arrives at the middle node at the exact same time as the quantum signal. This is possible since the speed of light through free space $c_\text{c}$ is faster than the speed of light through fibre $c_\text{q}$. We assume all operations, processing, switching, and measurements are infinitely fast (much faster than the light travel times between nodes).

We require that
\begin{align}
    t_2 &= t_1 + t_\text{c}.
\end{align}

We also have that 
\begin{align}
    L &= 2(2d_1 + d_2).
\end{align}

We also have the following times:
\begin{align}
    t_1 &= \frac{d_1}{c_\text{q}}, \; t_2 = \frac{d_2}{c_\text{q}}, \; t_\text{c} = \frac{d_\text{c}}{c_\text{c}}.
\end{align}

For fixed $L$, we can solve $d_2$ in terms of $d_1$:
\begin{align}
    d_2 &= \frac{L}{2} - 2d_1.
\end{align}

Then we can write
\begin{align}
    t_2 &= t_1 + t_\text{c}\\
    \frac{d_2}{c_\text{q}} &= \frac{d_1}{c_\text{q}} + \frac{d_\text{c}}{c_\text{c}}.
\end{align}

Also using
\begin{align}
    d_\text{c} &= d_1 + d_2,
\end{align}
to write
\begin{align}
    \frac{d_2}{c_\text{q}} &= \frac{d_1}{c_\text{q}} + \frac{d_1 + d_2}{c_\text{c}}.
\end{align}

Which gives
\begin{align}
    (c_\text{c} - c_\text{q}) d_2 &= (c_\text{c}+c_\text{q}) d_1.
\end{align}

Then,
\begin{align}
    (c_\text{c} - c_\text{q}) \left(\frac{L}{2} - 2d_1 \right) &= (c_\text{c}+c_\text{q}) d_1,
\end{align}
giving
\begin{align}
    d_1 &= \frac{L(c_\text{c} - c_\text{q})}{6 c_\text{c} - 2 c_\text{q}},\\
    d_2 &= \frac{L(c_\text{c} + c_\text{q})}{2(3c_\text{c} - c_\text{q})}.
\end{align}

Then we have
\begin{align}
    \eta_1 &= 10^{-\frac{L \alpha (c_\text{c} - c_\text{q})}{10 (6 c_\text{c} - 2 c_\text{q})}},\\
\eta_2 &= 10^{-\frac{L \alpha (c_\text{c} + c_\text{q})}{20 (3 c_\text{c} - c_\text{q})}}
\end{align}
and the key rate is
\begin{align}
    K_1 &= \frac{1}{2}P_1 P_2 = \frac{1}{2}\eta_1 \eta_2 = \frac{1}{2}10^{-\frac{L \alpha c_\text{c}}{30 c_\text{c} - 10 c_\text{q}}} = \frac{1}{2}\eta^{\frac{ c_\text{c}}{3 c_\text{c} - c_\text{q}}}.
\end{align}

Generally, we have $c_\text{q} \approx \frac{2}{3} c_\text{c}$, in which case we have
\begin{align}
    K_1 &= \frac{1}{2}10^{-\frac{3 L \alpha}{70}} = \frac{1}{2}\eta^{3/7}.
\end{align}

That is, our repeater does better than the total distance by a factor of $3/7$ which is better than a single repeater which can only improve to a factor of $1/2$ in th exponent. The optimal for three repeater nodes is $1/4$. So we have a $1/14$ improvement, and this is because $d_1/L = 1/14$, we have optimally ``removed'' a chunk of the channel.

As $c_\text{q}/c_\text{c} \to 0$, we have that 
\begin{align}
    K_{1, c_\text{q}/c_\text{c} \to 0} &= \frac{1}{2}\eta^{1/3},
\end{align}
which is the ultimate limit of our design with one level of nesting. To do better, we need more nesting levels.

\subsection{$N=2$}

Two levels of nesting is shown in the following circuit diagram:

\begin{figure}[H]
\centering
 \includegraphics[width=1\linewidth]{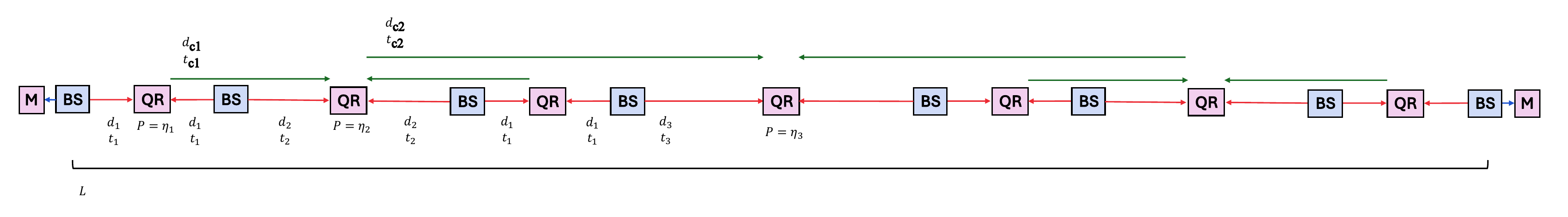}\label{fig:N3}
\end{figure}

Assuming $c_q = \frac{2}{3} c_c$ where $c_c = c$ where $c$ is the speed of light in vacuum, we obtain $K_2 = \frac{1}{2}\eta^{35/86}$.

\subsection{$N=3$}

Similarly, for $N=3$ we get $K_3 = \frac{1}{2}\eta^{104/259}$.

\subsection{$N=4$}

For $N=4$ we get $K_4 =\frac{1}{2} \eta^{249/622}$.

\subsection{$N=5$}

For $N=5$ we get $K_5 = \frac{1}{2}\eta^{3733/9331}$.

\subsection{$N \to \infty$}

For $N\to\infty$ we approach $K_\infty = \frac{1}{2}\eta^{2/5}$.

\subsection{General communication speeds $c_c$ and $c_q$}
For general $c_c$ and $c_q$, we calculate the rates as follows.
\begin{align*}
\text{scaling}_1 &= \frac{c_c}{3c_c - c_q} \\[8pt]
\text{scaling}_2 &= \frac{3c_c^2 + 2c_c c_q - c_q^2}{2(7c_c^2 - 4c_c c_q + c_q^2)} \\[8pt]
\text{scaling}_3 &= \frac{2c_c^3 + 5c_c^2 c_q - 4c_c c_q^2 + c_q^3}{15c_c^3 - 11c_c^2 c_q + 5c_c c_q^2 - c_q^3} \\[8pt]
\text{scaling}_4 &= \frac{5c_c^4 + 32c_c^3 c_q - 34c_c^2 c_q^2 + 16c_c c_q^3 - 3c_q^4}{2(31c_c^4 - 26c_c^3 c_q + 16c_c^2 c_q^2 - 6c_c c_q^3 + c_q^4)} \\[8pt]
\text{scaling}_5 &= \frac{3c_c^5 + 42c_c^4 c_q - 54c_c^3 c_q^2 + 36c_c^2 c_q^3 - 13c_c c_q^4 + 2c_q^5}{63c_c^5 - 57c_c^4 c_q + 42c_c^3 c_q^2 - 22c_c^2 c_q^3 + 7c_c c_q^4 - c_q^5}\\[8pt]
\text{scaling}_\infty &= \frac{c_q}{c_c+c_q}.
\end{align*}

Letting $c_c = 1$, $c_q = f c_c$ so that $f=c_q/c_c$, then we find:
\begin{align*}
\text{scaling}_1 &= \frac{-1}{f-3} \\[8pt]
\text{scaling}_2 &= \frac{- f^2 + 2f + 3}{2f^2 - 8f + 14} \\[8pt]
\text{scaling}_3 &= -\frac{f^3 - 4f^2 + 5f + 2}{f^3 - 5f^2 + 11f - 15} \\[8pt]
\text{scaling}_4 &= \frac{- 3f^4 + 16f^3 - 34f^2 + 32f + 5}{2f^4 - 12f^3 + 32f^2 - 52f + 62} \\[8pt]
\text{scaling}_5 &= -\frac{2f^5 - 13f^4 + 36f^3 - 54f^2 + 42f + 3}{f^5 - 7f^4 + 22f^3 - 42f^2 + 57f - 63} \\[8pt]
\text{scaling}_{\infty} &= \frac{f}{f+1}.
\end{align*}

The asymptotic rate is
\begin{align*}
K_{\infty} &= \frac{1}{2} \eta^{\text{scaling}_\infty}\\
&= \frac{1}{2}\eta^{\frac{c_q}{c_c + c_q}}.
\end{align*}

As $c_c >> c_q$, we find
\[
K_{\infty} = \frac{1}{2}.
\]
This means our scheme is scalable to arbitrarily long distances with high rates so long as the quantum signal can be arbitrarily slowed down compared to the classical signal.

We present a closed form for $\text{scaling}_N$ in the next section.

\section{Analytical limit of the nesting problem}

Let \(N\) be the nesting depth and define \(m=N+1\).
The decision variables are \(d_1,\dots,d_m\ge 0\).
Introduce the ratio
\[
f := \frac{c_q}{c_c}.
\]

\subsection{Constraints}

\paragraph{Total length.}
The total length constraint is
\[
L = 2\sum_{i=1}^{m} 2^{\,m-i} d_i .
\]
Define weighted partial sums
\[
S_k := \sum_{i=1}^{k} 2^{\,k-i} d_i , \qquad S_0 := 0 .
\]
Then the length constraint becomes
\[
S_m = \frac{L}{2}.
\]

\paragraph{Timing constraints.}
For \(k=1,\dots,N\), the timing constraints are
\[
\frac{d_{c_k}}{c_c} + \frac{d_k}{c_q} = \frac{d_{k+1}}{c_q},
\]
where
\[
d_{c_k} = \sum_{i=1}^{k} 2^{\,k-i} d_i + d_{k+1} = S_k + d_{k+1}.
\]
Multiplying by \(c_q\) and using \(f=c_q/c_c\) gives
\[
f (S_k + d_{k+1}) + d_k = d_{k+1},
\]
or equivalently
\begin{equation}
(1-f)d_{k+1} = d_k + f S_k .
\label{eq:dkrec}
\end{equation}

\subsection{Recurrence for the partial sums}

By definition,
\[
S_{k+1} = 2S_k + d_{k+1},
\qquad
d_k = S_k - 2S_{k-1}.
\]
Substituting into \eqref{eq:dkrec} yields
\[
(1-f)(S_{k+1}-2S_k) = (S_k-2S_{k-1}) + f S_k,
\]
which simplifies to the second-order linear recurrence
\begin{equation}
S_{k+1}
= \frac{3-f}{1-f} S_k
  - \frac{2}{1-f} S_{k-1}.
\label{eq:Skrec}
\end{equation}

\subsection{Characteristic roots}

The characteristic equation of \eqref{eq:Skrec} is
\[
\lambda^2 - \frac{3-f}{1-f}\lambda
+ \frac{2}{1-f} = 0,
\]
whose roots are
\[
\lambda_1 = 1,
\qquad
\lambda_2 = g := \frac{2}{1-f}
           = \frac{2c_c}{c_c-c_q}.
\]

\subsection{Feasibility}

Since \(d_k\ge 0\) and \(S_k\ge 0\), equation \eqref{eq:dkrec} implies
\[
(1-f)d_{k+1}\ge 0.
\]
For nontrivial solutions with \(L>0\), this requires
\[
f<1 \quad\Longleftrightarrow\quad c_q<c_c.
\]

\subsection{Closed form of \(S_k\)}

With roots \(1\) and \(g\) and condition \(S_0=0\), the solution is
\[
S_k = C(g^k - 1).
\]
Imposing \(S_m=L/2\) gives
\[
S_k = \frac{L}{2}\,
      \frac{g^k - 1}{g^m - 1}.
\]

\subsection{Objective function}

The objective is
\[
E_N = \sum_{k=1}^{m} d_k.
\]
We define the (dimensionless) scaling as
\[
\text{scaling}_N := \frac{E_N}{L}.
\]

Using \(d_k=S_k-2S_{k-1}\),
\[
E_N
= \sum_{k=1}^{m}(S_k-2S_{k-1})
= S_m - \sum_{k=1}^{m-1} S_k.
\]

\subsection{Limit as \(N\to\infty\)}

For \(c_q<c_c\), we have \(g>2\).
As \(m\to\infty\),
\[
\sum_{k=1}^{m-1} S_k
\sim S_m \sum_{t=1}^{\infty} g^{-t}
= \frac{S_m}{g-1}.
\]
Hence
\[
\lim_{N\to\infty} E_N
= S_m\left(1-\frac{1}{g-1}\right)
= S_m\frac{g-2}{g-1}.
\]
Substituting \(S_m=L/2\) and \(g=2/(1-f)\),
\[
\frac{g-2}{g-1}
= \frac{2f}{1+f}.
\]

\subsection{Final result}

Therefore, for \(c_q<c_c\),
\[
\boxed{
\lim_{N\to\infty} E_N
= L\,\frac{f}{1+f}
= L\,\frac{c_q}{c_c+c_q}.
}
\]

\[
\text{scaling}_N := \frac{E_N}{L}, 
\qquad m = N+1,
\qquad g = \frac{2}{1-f}.
\]

From the closed form of the partial sums,
\[
S_k = \frac{L}{2}\,\frac{g^k - 1}{g^{m} - 1},
\qquad
S_m = \frac{L}{2},
\]
and
\[
E_N = S_m - \sum_{k=1}^{m-1} S_k,
\]
we obtain
\[
\text{scaling}_N
= \frac{1}{2}
- \frac{1}{2}\sum_{k=1}^{m-1}
\frac{g^k - 1}{g^{m} - 1}.
\]

Evaluating the sums explicitly yields
\[
{
\text{scaling}_N(f)
=
\frac12
-
\frac12\,
\frac{
\displaystyle
\frac{g(g^{m-1}-1)}{g-1}
-(m-1)
}{
g^{m}-1
},
\qquad
g=\frac{2}{1-f},\ \ m=N+1.
}
\]

Moreover, the asymptotic scaling is
\[
\text{scaling}_\infty
= \lim_{N\to\infty}\text{scaling}_N
= \frac{f}{1+f}.
\]

The rate at nesting depth \(N\) is
\[
K_N := \frac{1}{2}\,\eta^{\text{scaling}_N},
\]
where \(\eta\) denotes the total channel transmissivity between the end nodes Alice and Bob.

In the asymptotic limit \(N\to\infty\), the rate converges to
\[
K_\infty
:= \lim_{N\to\infty} K_N
= \frac{1}{2}\,
\eta^{\text{scaling}_\infty}
= \frac{1}{2}\,
\eta^{\frac{f}{1+f}}
= \frac{1}{2}\,
\eta^{\frac{c_q}{c_c+c_q}}.
\]

\subsection{Quantum memory thresholds}

Consider storing the quantum states in quantum buffers after the QND measurement and before the BSM. One may ask whether such buffers can improve the quantum communication rate. We find that they provide no advantage until their loss rate falls below a critical threshold, which is typically on the order of half the loss rate of the quantum channel. Below this threshold, the buffers begin to function as genuine quantum memories and lead to a continuous improvement in performance. In the limit of vanishing memory loss, the protocol approaches the ideal repeater key-rate scaling. In the following, we analyse this behaviour explicitly for the $N=1$ protocol.

\paragraph{Definitions.}
For fixed \(L,c_q,c_c,c_\text{QM}\), where $L$ is the total distance and $c_q,c_c,c_\text{QM}$ are the speeds of the quantum, classical, and quantum memory buffer channels, respectively, let
\begin{equation}
f \;\equiv\; \frac{c_q}{c_c},
\qquad
\gamma \;\equiv\; \frac{\alpha_{\mathrm{QM}}}{\alpha}.
\end{equation}
The scaling cost is
\begin{equation}
E_1
= d_1 + d_2 + \gamma(s_1+s_2),
\end{equation}
subject to
\begin{equation}
L = 2(2d_1+d_2),
\end{equation}
and the timing constraint
\begin{equation}
\frac{d_1+d_2}{c_c} + \frac{d_1}{c_q} + \frac{s_1}{c_\text{QM}}
=
\frac{d_2}{c_q} + \frac{s_2}{c_\text{QM}},
\end{equation}
with all variables non-negative.

\paragraph{Critical point.}
The minimisation exhibits a critical value \(\gamma^*\) above which the
optimal solution has no buffering (\(s_1=s_2=0\)). The threshold is
\begin{equation}
\boxed{
\gamma^*
= \frac{c_q c_c}{c_\text{QM}(3c_c-c_q)}
= \frac{f\,c_c}{c_\text{QM}(3-f)}
},
\qquad f<3 .
\end{equation}

\paragraph{Scaling below the critical point.}
For \(0 \le \gamma \le \gamma^*\) (and in particular for \(f \le 1\)), the optimum occurs at
\(
d_1=L/4,\; d_2=0
\),
and the minimum scaling is linear in \(\gamma\):
\begin{equation}
\boxed{
E_{1,\min}(\gamma)
=
\frac{L}{4}
+\gamma\,\frac{c_\text{QM} L}{4}
\left(\frac{1}{c_q}+\frac{1}{c_c}\right)
=
\frac{L}{4}
+\gamma\,\frac{c_\text{QM} L}{4c_c}\!\left(1+\frac{1}{f}\right)
}.
\end{equation}

\paragraph{Scaling above the critical point.}
For \(\gamma \ge \gamma^*\), buffering is suppressed,
\(s_1=s_2=0\), and the minimum scaling is independent of \(\gamma\):
\begin{equation}
\boxed{
E_{1,\min}
= d_1+d_2
= \frac{L}{3-f}
}.
\end{equation}

Equivalently, the result can be summarised as the piecewise form
\begin{equation}
\boxed{
E_{1,\min}(\gamma)
=
\begin{cases}
\displaystyle
\frac{L}{4}
+\gamma\,\frac{c_\text{QM} L}{4c_c}\!\left(1+\frac{1}{f}\right),
& 0 \le \gamma \le \gamma^*, \\[1.2em]
\displaystyle
\frac{L}{3-f},
& \gamma \ge \gamma^* .
\end{cases}
}
\end{equation}

\paragraph{Concrete parameter regimes.}
We now specialise to the physically relevant case
\(
c_c = 1,\; c_\text{QM} = 1,\; L = 1,
\)
corresponding to classical signalling and quantum memory buffers operating at (approximately) the speed of light in air, over a unit total distance. We consider two representative quantum-channel speeds: \(c_q = 1\) and \(c_q = 2/3\), corresponding to \(f = 1\) and \(f = 2/3\), respectively. 

For \(c_q = 1\), the critical memory-quality threshold is
\(
\gamma^* = \tfrac{1}{2},
\)
and the minimum scaling is
\[
E_{1,\min}(\gamma) =
\begin{cases}
\frac{1}{2}, & \gamma \ge \frac{1}{2}, \\[6pt]
\frac{1}{4} + \frac{\gamma}{2}, & \gamma < \frac{1}{2}.
\end{cases}
\]
Using \(\gamma = \alpha_{\mathrm{QM}}/\alpha\) and fixing a standard fibre loss
\(\alpha = 0.2~\mathrm{dB/km}\), this corresponds to a critical quantum-memory loss rate
\[
\alpha_{\mathrm{QM}}^{\mathrm{crit}} = 0.1~\mathrm{dB/km}.
\]

For \(c_q = 2/3\), the threshold shifts to
\(
\gamma^* = \tfrac{2}{7},
\)
with
\[
E_{1,\min}(\gamma) =
\begin{cases}
\frac{3}{7}, & \gamma \ge \frac{2}{7}, \\[6pt]
\frac{1}{4} + \frac{7\gamma}{8}, & \gamma < \frac{2}{7}.
\end{cases}
\]
In this case, the corresponding memory-loss threshold is
\[
\alpha_{\mathrm{QM}}^{\mathrm{crit}}
= \frac{2}{7}\,\alpha
\approx 0.057~\mathrm{dB/km}.
\]

In both regimes, quantum buffering provides no advantage unless the memory loss rate is below a well-defined threshold set by the relative quantum and classical communication speeds. Once this threshold is crossed, buffering acts as a genuine quantum memory and leads to a continuous improvement in the scaling. In the limit of vanishing memory loss, the protocol approaches the ideal repeater scaling.

Finally, we note that current quantum buffer implementations may have loss rates above \(0.1~\mathrm{dB/km}\). In this regime, quantum buffering does not provide a benefit, and the optimal strategy is therefore to operate without long-lived quantum memories. Importantly, however, the fast classical communication assumed in our protocol (\(c_c \simeq c_\text{QM} \simeq c > c_q\)) is sufficient to surpass the single-repeater bound even with existing technology. This demonstrates that, in contrast to memory-based repeater architectures, improvements in classical communication speed alone can already enable performance beyond the single-repeater limit, without requiring high-quality quantum memories.

\section{Secret key rate derivation of our practical protocol}\label{sec:SKR_derivation}

We aim to calculate a lower bound on the achievable key rate for our practical multi-node protocol based on single-photon interference, as shown in Fig.~2(a) of the main text.

For our protocol, the secret key rate (SKR) is given by:
\begin{align}
\text{SKR} = \frac{r R}{\tau},
\end{align}
where $r$ is the raw key rate obtainable from Alice and Bob's data upon success of the repeater protocol, $R$ is the success rate of the entire repeater protocol, and $\tau = 1/R_\text{rep.}$ is the time in seconds for each pulse, the inverse of the repetition rate $R_\text{rep.}$.

We can lower bound $r$ using the (reverse) coherent information, $I$~\cite{Garc_a_Patr_n_2009}.

Let's consider an ideal implementation of our quantum repeater protocol for pure-loss quantum channels. Then, the optimal raw key rate $r$ is equal to one bit per use of the channel in the limit of small mean photon numbers prepared at Alice and at Bob, which is proved in the state derication section.

The repeater rate $R$ for our setup with four links is equal to:
\begin{align}
    R &= \frac{1}{\overline{Z}_0^{(m)}(P_0)} \frac{1}{\overline{Z}_1 \left( P_1 \right)},
\end{align}
where $P_0$ is the probability that the two Bell-state measurements are successful in the memory-less nodes and $P_1$ is the probability that the one Bell-state measurement is successful in the quantum memory node. ${\overline{Z}_1 \left( P_1 \right)}$ and ${\overline{Z}_0^{(m)}(P_0)}$ are the wait times for successful events in number of steps, as we will discuss in a moment.

Let us define the following parameters:

\begin{itemize}
    \item \( d_1, d_2, d_{\text{QM}_1}, d_{\text{QM}_2} \): Distances for different segments as shown in the figure.
    \item \( m \): maximum number of steps waiting in the memory ($d_{\text{QM}_2}$).
    \item \( \tau \): average time to send each pulse (inverse of the repetition rate).
    \item \( \alpha, \alpha_{\text{QM}} \): Channel loss rate for the quantum channel and quantum memory channel, respectively.
    \item \( \chi \): TMSV state of Alice and Bob in the entanglement-based version of the protocol.
    \item \( L \): Total distance.
    \item \( c_q, c_c, c_{\text{QM}} \): Speeds of light in quantum channels, classical channels, and quantum memory channels, respectively.
\end{itemize}

Step 1: Solving for \( d_1 \)

Given the total length \( L \), we have:
\[
L = 2 \times (2 d_1 + d_2).
\]

Solving for \( d_1 \), we get:
\[
d_1 = \frac{L}{4} - \frac{d_2}{2}.
\]

Step 2: Calculating Time Delays

Let us calculate the time delays for each segment.

1. Time delay for \( d_1 \) through quantum channel:
   \[
   t_1 = \frac{d_1}{c_q}
   \]

2. Time delay for \( d_2 \) through quantum channel:
   \[
   t_2 = \frac{d_2}{c_q}
   \]

3. Time delay for \( d_{\text{QM}_1} \) through quantum memory:
   \[
   t_{\text{QM}_1} = \frac{d_{\text{QM}_1}}{c_{\text{QM}}}
   \]

4. Time delay for classical communication through \( d_1 + d_2 \):
   \[
   t_c = \frac{d_1 + d_2}{c_c}
   \]

Setting \( t_{\text{QM}_1} = \text{max}(t_1 + t_c - t_2,0) \), we solve for \( d_{\text{QM}_1} \):
\[
d_{\text{QM}_1} = \frac{L \, c_{\text{QM}} \, c_c + L \, c_{\text{QM}} \, c_q - 6 \, c_{\text{QM}} \, c_c \, d_2 + 2 \, c_{\text{QM}} \, c_q \, d_2}{4 \, c_c \, c_q}
\]

At this point, we impose a condition on \( d_2 \) such that \( 0 < d_2 < \frac{ L (c_c +  c_q)}{6 \, c_c - 2 \, c_q} \). This constraint prevents \( t_{\text{QM}_1} \) from becoming ``negative,'' which would physically imply that the classical communication arrives before the quantum signal. In such a case, the setup would be inefficient, as the quantum signal would waste time in the channel, accumulating unnecessary losses. We are not interested in configurations where the classical signal arrives first.

Step 3: Solving for \( d_{\text{QM}_2} \)

The time delay for \( d_{\text{QM}_2} \) through the quantum memory is given by:
\[
t_{\text{QM}_2} = \frac{d_{\text{QM}_2}}{c_{\text{QM}}}
\]

Setting \( t_{\text{QM}_2} = m \tau \), we solve for \( d_{\text{QM}_2} \):
\[
d_{\text{QM}_2} = c_{\text{QM}} \, m \, \tau
\]

Note that $d_{\text{QM}_2}$ is only required on one side. We assume it is on both sides since this is a lower bound on the rate and in practice more loss could be added (e.g., with a variable beamsplitter) to balance both sides.

Step 4: Transmission Efficiencies

The transmissivity \( \eta \) over distance \( L \) (km) with loss rate \( \alpha \) (dB/km) is given by
\[
\eta = 10^{-\alpha L / 10}.
\]

The transmission efficiencies \( \eta_i \) for each segment are given by:

1. Transmission efficiency for \( d_1 \):
   \[
   \eta_1 = 10^{-\frac{\alpha}{10} \cdot d_1}
   \]

2. Transmission efficiency for \( d_2 \):
   \[
   \eta_2 = 10^{-\frac{\alpha}{10} \cdot d_2}
   \]

3. Transmission efficiency for \( d_{\text{QM}_1} \):
   \[
   \eta_{\text{QM}_1} = 10^{-\frac{\alpha_{\text{QM}}}{10} \cdot d_{\text{QM}_1}}
   \]

4. Transmission efficiency for \( d_{\text{QM}_2} \):
   \[
   \eta_{\text{QM}_2} = 10^{-\frac{\alpha_{\text{QM}}}{10} \cdot d_{\text{QM}_2}}
   \]

Step 5: Calculating Probabilities

Define the probabilities in the limit of small $\chi$ as follows:

1. Probability \( P_0 \):
   \[
   P_0 = \frac{\eta_1}{2}
   \]

2. Probability \( P_1 \):
   \[
   P_1 = 2 \, \chi^2 \, \eta_2 \, \eta_{\text{QM}_1} \, \eta_{\text{QM}_2} \, \eta_\text{switch}^{b \lceil \log_b(m) \rceil + 1}
   \]

We also define \( q \):
   \[
   q = 1 - P_0
   \]

Step 6: Defining \( Z_0 \) and \( Z_1 \)

1. Expression for \( Z_0 \)~\cite{PhysRevA.100.032322}:
   \[
   Z_0 = \frac{1 + 2q - 2q^{m+1}}{P_0 \, (1 + q - 2q^{m+1})}
   \]

2. Expression for \( Z_1 \):
   \[
   Z_1 = \frac{1}{P_1}
   \]

Step 7: Key Rate \( K \)

The key rate \( K \) is given by:
\[
K = \frac{1}{Z_0 \, Z_1 \, \tau}
\]
After simplification, in the limit of small $\chi$, \( K \) becomes:
\[
K = \frac{\chi^2 \, \eta_1 \, \eta_2 \, \eta_{\text{QM}_1} \, \eta_{\text{QM}_2} \, \eta_\text{switch}^{b \lceil \log_b(m) \rceil + 1} \left( \frac{\eta_1}{2} + 2 \left( 1 - \frac{\eta_1}{2} \right)^{m + 1} - 2 \right)}{\tau \left( \eta_1 + 2 \left( 1 - \frac{\eta_1}{2} \right)^{m + 1} - 3 \right)} \text{ bits/s},
\]

This completes the derivation of the analytical key rate.

The key rate \( K \) depends on the following parameters:

\begin{itemize}
    \item \( \chi \): Two-mode squeezed vacuum (TMSV) parameter at Alice and Bob. A higher \( \chi \) (more photons) can increase the rate; however, too many photons may saturate the noiseless linear amplifier, degrading the rate. The approximate analytical rate we express is correct in the limit of small \( \chi \).
    \item \( \alpha \): Channel loss rate. Lower channel loss increases the rate by preserving more photons through the channel.
    \item \( \alpha_{\text{QM}} \): Quantum memory loss rate. A smaller \( \alpha_{\text{QM}} \) improves the rate and allows scaling close to the three-repeater bound.
    \item \( L \): Total transmission distance. Combined with \( d_2 \), it determines the positions of all nodes, optimising transmission.
    \item \( d_2 \): Intermediate distance parameter. This sets the node positions in relation to \( L \) and determines \( d_1 \); \( d_2 \) should be chosen to maximise the rate.
    \item \( m \): Maximum number of storage steps in the fibre loop of length \( d_{\text{QM}_2} \). The rate can be optimised by adjusting \( m \).
    \item \( \tau \): Inverse of the repetition rate. Higher repetition rates mean photons are stored for shorter times, thereby improving the rate.
    \item \( c_{\text{QM}} \): Speed of light in the quantum memory medium. Slower speeds in the quantum memory mean photons can be stored longer, which improves the rate.
    \item \( c_c \): Speed of light in the classical communication channel.
    \item \( c_q \): Speed of light in the quantum communication channel. A slower quantum channel speed allows classical signals to reach nodes sooner, effectively increasing the rate by reducing the storage time.
\end{itemize}

The experimenter has flexibility to adjust \( d_2 \), \( m \), and \( \chi \) to optimise the key rate \( K \), while all other parameters are typically fixed by experimental constraints.

Specifically:

- Adjusting \( d_2 \): The parameter \( d_2 \) sets the position of intermediate nodes within the total transmission distance \( L \). By varying \( d_2 \), the experimenter can adjust the relative positions of all nodes, including the effective lengths \( d_1 \) and \( d_{\text{QM}_1} \), to optimise the key rate by balancing the distances travelled by quantum and classical signals.

- Adjusting \( m \): The parameter \( m \) represents the maximum number of storage steps within a fibre loop of length \( d_{\text{QM}_2} \), controlling the time quantum signals are stored in the memory.

- Optimising \( \chi \): The parameter \( \chi \) is the two-mode squeezed vacuum (TMSV) parameter, governing the mean number of photons Alice and Bob send. While a higher \( \chi \) can increase the key rate by sending more photons, too many photons can saturate the noiseless linear amplifier, leading to degradation in the rate. Thus, \( \chi \) should be carefully optimised to balance these effects. However, the analytical expression for the key rate is only valid within the limit of small \( \chi \).

All other parameters—such as \( \alpha \), \( \alpha_{\text{QM}} \), \( L \), \( \tau \), \( c_{\text{QM}} \), \( c_c \), and \( c_q \)—are usually constrained by the experimental setup, material properties, and operational limits. Therefore, \( d_2 \), \( m \), and \( \chi \) provide the primary levers for optimising system performance within these constraints. 

In practice, distances between nodes may vary; however, we focus here on the balanced setup (just $d_1$ and $d_2$). Additional losses can be introduced to equalise the distances, or the beamsplitter ratios and \( \chi \) can be adjusted to optimise the rate for any given set of distances.

\section{State and success probabilities derivation}\label{sec:state_derivation}

In this section, we analytically derive the global output state of our practical multi-node protocol based on single-photon interference, as shown in Fig.~2(a) of the main text step by step. While this process is significantly easier to perform numerically, we provide analytical results to highlight that, in the entanglement-based (EB) version for pure-loss channels, the final output state shared by the trusted parties remains pure and maximally entangled in the limit of small photon numbers prepared by the trusted parties.

\subsection{Roadmap}

Below, we outline the roadmap for this calculation, which follows the same structure whether performed numerically or analytically:
\begin{figure}[H]
    \centering
    \includegraphics[scale=0.7]{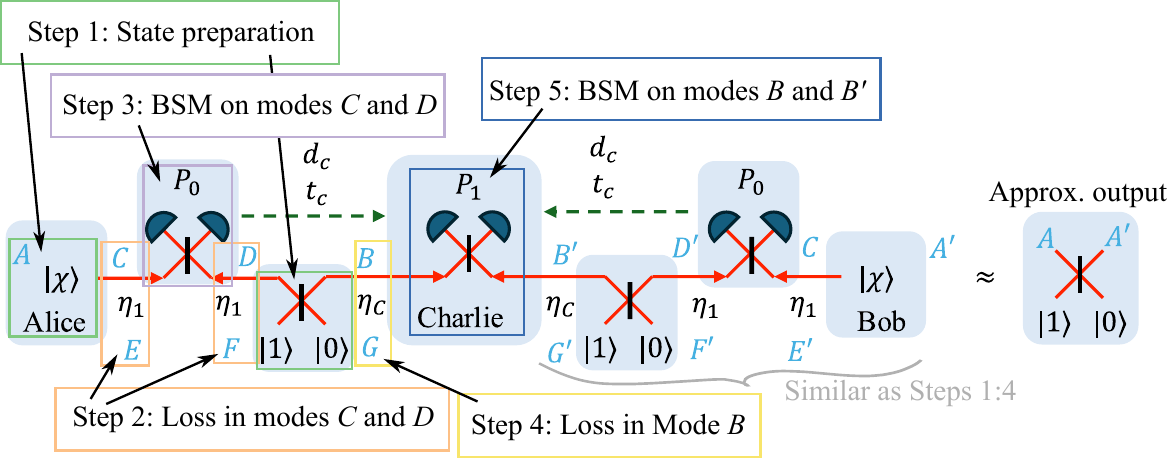}
    \label{fig:der_1}
\end{figure}

In Step 1, we prepare Alice's initial state of modes \( A \) and \( C \), denoted as \( \ket{\chi}_{A,C} \), and the relay node's entanglement resource, which consists of a single photon \( \ket{1}_B \) and vacuum \( \ket{0}_D \) mixed on a beamsplitter with transmissivity \( 1/2 \). 

In Step 2, we introduce loss in modes \( C \) and \( D \) by coupling them to vacuum environmental modes \( E \) and \( F \), respectively, via beamsplitters with transmissivity \( \eta_1 \). (For brevity, these beamsplitters are not shown in the diagram.)  

In Step 3, we perform a Bell-state measurement (BSM) on modes \( C \) and \( D \), which involves interfering them on a beamsplitter with transmissivity \( 1/2 \) and detecting a single photon at one detector and vacuum at the other detector. This process effectively teleports the state of mode \( C \) to mode \( B \) with high fidelity and high rate (scaling optimally with loss). In the limit of small \( \chi \), the success probability is given by  
\[
P_0 \approx \frac{\eta_1}{\sqrt{2}},
\]
\noindent where a phase correction may be applied if necessary or alternatively tracked in software, depending on which detector registers the photon.

In Step 4, we apply loss to mode \( B \) by coupling it to a vacuum environmental mode \( G \) with transmissivity $\eta_C$. We then prepare two copies of the state derived in Steps 1–4, with modes \( B \) and \( B' \) stored in lossy quantum memories at Charlie until they are ready. $\eta_C$ includes transmission losses and memory losses.

In Step 5, we perform a BSM on modes \( B \) and \( B' \) with success probability  
\[
P_1 \approx 2\chi^2 \eta_C.
\]

This total process results in the final renormalised entangled state  
\[
\ket{\psi}_{A,A'} \approx \frac{\ket{1,0} + \ket{0,1}}{\sqrt{2}},
\]  
which contains one ebit of entanglement. 

The rate \( R \) at which successful entanglement is generated depends on the success probabilities \( P_0 \) and \( P_1 \), as well as the repeater operation, specifically the switching and matching procedures, which is derived in another section.

In the following derivations, each step involves equations accompanied by figures to illustrate the physical setup and transformations in a clear way. We consider the global pure state throughout the calculation. That is, loss is modelled by introducing environment vacuum modes and mixing with a beamsplitter with the transmissivity of the lossy channel. Measurements is a projection onto a Bell state. Finally, we derive the output state of the entire repeater protocol and the success probabilities. Key equations and probability expressions will be provided along the way.

For notational convenience, we use multiple equivalent representations for Fock states across different modes, depending on the context and level of brevity required:  
\[
|n\rangle_A \otimes |m\rangle_B \otimes \dots = |n\rangle_A |m\rangle_B \dots = |n,m, \dots\rangle_{A,B,\dots} = |n,m, \dots\rangle.
\]

\subsection{Beamsplitter}

A beamsplitter acts as a unitary transformation on the mode operators:
\begin{equation}
    \begin{pmatrix} \hat{a}_1' \\ \hat{a}_2' \end{pmatrix} = 
    \begin{pmatrix} t & r \\ -r^* & t^* \end{pmatrix} 
    \begin{pmatrix} \hat{a}_1 \\ \hat{a}_2 \end{pmatrix},
\end{equation}
where $t$ and $r$ are the transmission and reflection coefficients satisfying $|t|^2 + |r|^2 = 1$.

Consider an initial Fock state $|n,0\rangle$:
\begin{equation}
    |n,0\rangle = \frac{(\hat{a}_1^\dagger)^n}{\sqrt{n!}} |0,0\rangle.
\end{equation}

The transformation of the creation operator is
\begin{equation}
    (\hat{a}_1')^\dagger = t^* \hat{a}_1^\dagger + r^* \hat{a}_2^\dagger.
\end{equation}
Applying this $n$ times,
\begin{equation}
    (\hat{a}_1')^{\dagger n} = (t^* \hat{a}_1^\dagger + r^* \hat{a}_2^\dagger)^n.
\end{equation}
Expanding via the binomial theorem,
\begin{equation}
    (\hat{a}_1')^{\dagger n} = \sum_{k=0}^{n} \binom{n}{k} (t^*)^{n-k} (r^*)^k (\hat{a}_1^\dagger)^{n-k} (\hat{a}_2^\dagger)^k.
\end{equation}

Applying to the vacuum state,
\begin{equation}
    B |n,0\rangle = \sum_{k=0}^{n} \binom{n}{k} (t^*)^{n-k} (r^*)^k |n-k, k\rangle.
\end{equation}
This shows that a beamsplitter distributes the initial photons binomially across the two output modes.

\subsection{Initial State Preparation}

In the entanglement-based (EB) version of the protocol, Alice prepares the initial state:

\begin{figure}[H]
    \centering
    \includegraphics[scale=0.7]{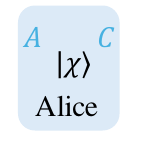}
    \label{fig:der_1a}
\end{figure}

The state is given by:
\begin{equation}
    \ket{\chi}_{AC} {=}  \sqrt{1{-}\chi^2} \sum_{n=0}^\infty \chi^n \ket{n}_A\ket{n}_C,
\end{equation}
with two-mode-squeezing parameter $r$, variance $\nu{=}\cosh{2r}$, $\chi{=}\tanh{r}$, and mean photon number $\bar{n}{=}\sinh^2{r}$, and where $\ket{n}$ are Fock-number states.

\subsection{Beamsplitter Transformation at a Node}

At one of the nodes, we introduce a single photon on a beamsplitter:

\begin{figure}[H]
    \centering
    \includegraphics[scale=0.7]{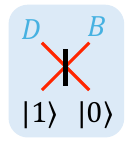}
    \label{fig:der_2}
\end{figure}

The corresponding quantum state is:
\begin{align}
    |\psi \rangle_{DB} &= \frac{  \ket{0}_D\ket{1}_B +  \ket{1}_D\ket{0}_B}{\sqrt{2}}.
\end{align}

\subsection{Loss Modelling in Modes C and D}

To model loss in modes C and D, we introduce auxiliary modes E and F, which are combined on a beamsplitter with transmissivity $\eta_1$. This is shown in the following circuit, where we omit beam splitters interacting with the environment and signal modes for brevity.

\begin{figure}[H]
    \centering
    \includegraphics[scale=0.7]{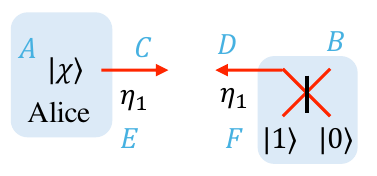}
    \label{fig:der_3}
\end{figure}

The resulting state is:
\begin{multline}
|\psi\rangle_{ACEDBF} \to  \sqrt{1{-}\chi^2} \sum_{n=0}^\infty \chi^n \sum_{k=0}^n \sqrt{{n \choose k}}
(1{-}\eta_1)^{\frac{k}{2}} \eta_1^{\frac{n-k}{2}}|n\rangle_{A} |n-k\rangle_{C} |k\rangle_E\\
 \biggl[ \frac{1}{\sqrt{2}} \ket{0}_D\ket{1}_B \ket{0}_F +  \frac{1}{\sqrt{2}} \bigl ( \sqrt{\eta_1} \ket{1}_D\ket{0}_B  \ket{0}_F + \sqrt{1{-}\eta_1}\ket{0}_D\ket{0}_B \ket{1}_F \bigl )\biggr].
\end{multline}

\subsection{Bell State Measurement (BSM)}

A Bell state measurement (BSM) is performed on modes C and D as illustrated:

\begin{figure}[H]
    \centering
    \includegraphics[scale=0.7]{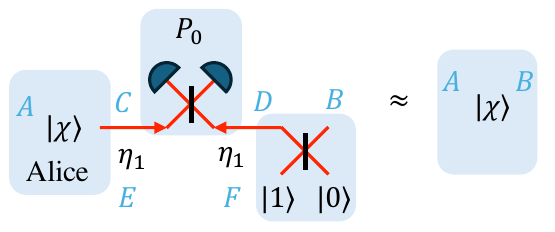}
    \label{fig:der_4}
\end{figure}

The Bell-state measurement (BSM) corresponds to a projection onto the Bell state
\[
|\Psi^\pm\rangle = \frac{1}{\sqrt{2}} \left( |01\rangle \pm |10\rangle \right).
\]
The corresponding projector is given by
\[
P_{\Psi^\pm} = |\Psi^\pm\rangle \langle \Psi^\pm|.
\]

Applying this projector to a state \( \ket{\psi} \) (unnormalised),
\[
\ket{\psi} \to P_{\Psi^\pm} \ket{\psi}.
\]

The probability is given by the Born rule:
\[
P = \bra{\psi} P_{\Psi^\pm} \ket{\psi} = \| P_{\Psi^\pm} \ket{\psi} \|^2.
\]

Upon a successful measurement, the state collapses to
\[
\ket{\psi} \to \frac{P_{\Psi^\pm} \ket{\psi}}{\sqrt{P}}.
\]

Since there are two successful outcomes (one for \( |\Psi^+\rangle \) and one for \( |\Psi^-\rangle \)), the total success probability is
\[
P_0 = 2P.
\]

The unnormalised output state after the BSM on modes C and D is:
\begin{multline}
|\psi\rangle_{ABEF} = \sqrt{\frac{1{-}\chi^2}{2}} {\eta}^\frac{1}{4} \sum_{k=0}^\infty  (1{-}\sqrt{\eta})^\frac{k}{2} 
\biggl[    \chi^k    \ket{k}_A\ket{0}_B\ket{k}_E\ket{0}_F \\ +   \chi^{k{+}1} \sqrt{k{+}1}  \bigl( \ket{k{+}1}_A\ket{1}_B\ket{k}_E\ket{0}_F + \sqrt{1{-}\sqrt{\eta}} \ket{k{+}1}_A\ket{0}_B\ket{k}_E\ket{1}_F \bigr)  \biggr],\label{eq:tele}
\end{multline}
which after normalisation is approximately equal to $\ket{\chi}_{AB}$ for small $\chi$, which means Alice's mode C has effectively been teleported across the relay to mode B.

The probability is
\[
P = \| \ket{\psi}_{ABEF} \|^2.
\]

The normalised state is
\[
\ket{\psi}_{ABEF} \to \frac{\ket{\psi}_{ABEF}}{\sqrt{P}},
\]
where the normalised state is $\approx \ket{\chi}$ for small $\chi$.

The total probability of success for one or the other successful outcome at this relay is:
\begin{equation}
    P_0 = 2P,
\end{equation}
where $P_0 \approx \frac{\eta_1}{\sqrt{2}}$ for small $\chi$.

\subsection{Further Loss in Mode B}

To model loss in mode B with transmissivity $\eta_C$, we introduce an auxiliary mode G:

\begin{figure}[H]
    \centering
    \includegraphics[scale=0.7]{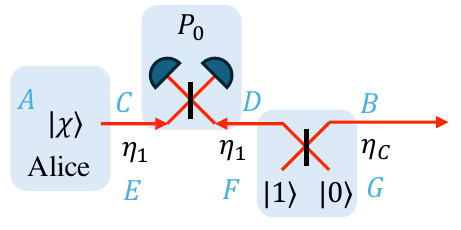}
    \label{fig:der_5}
\end{figure}

Note that $\eta_C$ captures the channel loss over distance  $d_2$ as well as the delay lines $d_{\text{QM}_1}$, and $d_{\text{QM}_2}$, the switching losses, and the variable optical buffer losses. That is, the worst-case transmissivity is given by $ \eta_C = \eta_2 \, \eta_{\text{QM}_1}  \, \eta_{\text{QM}_2} \, \eta_{\text{switch}}^{b \lceil \log_b(m) \rceil + 1}$, where $b$ is the base of the digital quantum memory and $m$ is the fixed number of storage cycles  ($t_{\text{QM}_2} = m\tau$), where $R_{\text{rep.}} = 1/\tau$.

The resulting normalised state is:

\[
\ket{\psi}_{A,E,B,F,G} = \frac{1}{2} \frac{1}{\sqrt{P}}\sum_{k=1}^{3} \ket{\psi}_k
\]

where the components \(\ket{\psi}_k\) are defined as follows:

\[
\ket{\psi}_1 = \sum_{n=0}^{\infty} { \chi^n \eta^{\frac{1}{2}} (1 - \chi^2)^{\frac{1}{2}} (1 - \eta)^{\frac{n}{2}}}
 \, \ket{n, n, 0, 0, 0}
\]

\[
\ket{\psi}_2 = \sum_{n=1}^{\infty} \sum_{j=0}^{1} { \chi^n \eta^{\frac{1}{2}} \eta_2^{\frac{1}{2} - \frac{j}{2}} n^{\frac{1}{2}} (1 - \chi^2)^{\frac{1}{2}} (1 - \eta)^{\frac{n}{2} - \frac{1}{2}} (1 - \eta_2)^{\frac{j}{2}}}
 \, \ket{n,n-1,1-j,0,j}
\]

\[
\ket{\psi}_3 = \sum_{n=1}^{\infty} { \chi^n \eta^{\frac{1}{2}} n^{\frac{1}{2}} (1 - \chi^2)^{\frac{1}{2}} (1 - \eta)^{\frac{1}{2}} (1 - \eta)^{\frac{n}{2} - \frac{1}{2}}}
 \, \ket{n,n-1,0,1,0}.
\]

\subsection{Final Bell State Measurement}

Finally, we perform a Bell state measurement on two copies of the above state:

\begin{figure}[H]
    \centering
    \includegraphics[scale=0.7]{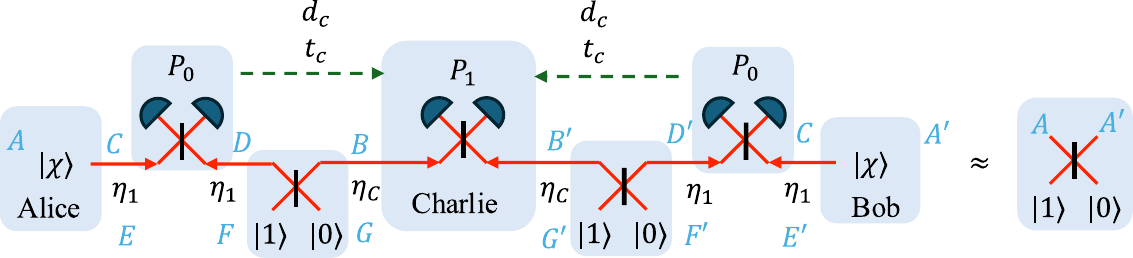}
    \label{fig:der_6}
\end{figure}

The final unnormalised global output state after Charlie's BSM on modes B and B' (remembering that modes $C$, $D$, $C'$, $D'$ have already been measured earlier) shared between Alice (\(A\)) and Bob (\(A'\)) and the six environment modes (\(E,F,G,E',F',G'\)) is

\[
|\psi\rangle_{A, E, F, G, A', E', F', G'} = \frac{1}{4} \frac{1}{P}   \sum_{k=1}^{6} |\psi_k\rangle_{A, E, F, G, A', E', F', G'}.
\]

We omit the mode labelling from now on to keep things concise. We define $|\psi_k\rangle$ as follows.

The component \( |\psi_1\rangle \) is given by:

\[
|\psi_1\rangle = \sum_{n=0}^{\infty} \sum_{m=1}^{\infty} c_1 c_2 \, |n,n,0,0,m,m-1,0,0\rangle,
\]

\noindent with the coefficients defined as:

\[
c_1 =  \, \chi^n \, \eta^{1/2} \, (1 - \chi^2)^{1/2} \, (1 - \eta)^{n/2},
\]

\[
c_2 =  \chi^n \sqrt{\eta} \sqrt{\eta_2} \sqrt{n} \sqrt{1 - \chi^2} (1 - \eta)^{n/2 - 1/2},
\]

where \( |n\rangle \) and \( |m\rangle \) denote the Fock states.

The component \(|\psi_2\rangle \) is given by:

\[
|\psi_2\rangle = \sum_{n=1}^{\infty} \sum_{m=0}^{\infty} c_3 c_4 \, |n,n-1,0,0,m,m,0,0\rangle,
\]

\noindent with the coefficients:

\[
c_3 =  \chi^n \sqrt{\eta} \sqrt{\eta_2} \sqrt{n} \sqrt{1 - \chi^2} (1 - \eta)^{n/2 - 1/2},
\]

\[
c_4 =  \, \chi^m \, \eta^{1/2} \, (1 - \chi^2)^{1/2} \, (1 - \eta)^{m/2}.
\]

The component \(|\psi_3\rangle \) is given by:

\[
|\psi_3\rangle = \sum_{n=1}^{\infty} \sum_{m=1}^{\infty} c_5 c_6 \, |n,n-1,0,1,m,m-1,0,0\rangle,
\]

\noindent with the coefficients:

\[
c_5 =  \, \chi^n \, \eta^{1/2} \, \sqrt{n} \, (1 - \chi^2)^{1/2} \, (1 - \eta)^{n/2 - 1/2} \, (1 - \eta_C)^{1/2},
\]

\[
c_6 =  \, \chi^m \, \eta^{1/2} \, \eta_C^{1/2} \, \sqrt{m} \, (1 - \chi^2)^{1/2} \, (1 - \eta)^{m/2 - 1/2}.
\]

The component \(|\psi_4\rangle \) is given by:

\[
|\psi_4\rangle = \sum_{n=1}^{\infty} \sum_{m=1}^{\infty} c_7 c_8 \, |n,n-1,0,0,m,m-1,0,1\rangle,
\]

\noindent with the coefficients:

\[
c_7 =  \, \chi^n \, \eta^{1/2} \, \eta_C^{1/2} \, \sqrt{n} \, (1 - \chi^2)^{1/2} \, (1 - \eta)^{n/2 - 1/2},
\]

\[
c_8 =  \, \chi^m \, \eta^{1/2} \, \sqrt{m} \, (1 - \chi^2)^{1/2} \, (1 - \eta)^{m/2 - 1/2} \, (1 - \eta_C)^{1/2}.
\]

The component \(|\psi_5\rangle \) is given by:
\[
|\psi_5\rangle = \sum_{n=1}^{\infty} \sum_{m=1}^{\infty} c_9 c_{10} \, |n,n-1,0,0,m,m-1,1,0\rangle,
\]

\noindent with the coefficients:

\[
c_9 =  \, \chi^n \, \eta^{1/2} \, \eta_C^{1/2} \, \sqrt{n} \, (1 - \chi^2)^{1/2} \, (1 - \eta)^{n/2 - 1/2},
\]

\[
c_{10} =  \, \chi^m \, \eta^{1/2} \, \sqrt{m} \, (1 - \chi^2)^{1/2} \, (1 - \eta)^{1/2} \, (1 - \eta)^{m/2 - 1/2}.
\]

The component \(|\psi_6\rangle \) is given by:

\[
|\psi_6\rangle = \sum_{n=1}^{\infty} \sum_{m=1}^{\infty} c_{11} c_{12} \, |n,n-1,1,0,m,m-1,0,0\rangle,
\]

\noindent with the coefficients:

\[
c_{11} =  \, \chi^n \, \eta^{1/2} \, \sqrt{n} \, (1 - \chi^2)^{1/2} \, (1 - \eta)^{1/2} \, (1 - \eta)^{n/2 - 1/2},
\]

\[
c_{12} =  \, \chi^m \, \eta^{1/2} \, \eta_C^{1/2} \, \sqrt{m} \, (1 - \chi^2)^{1/2} \, (1 - \eta)^{m/2 - 1/2}.
\]

\noindent The unormalised density matrix is then:

\[
\rho = |\psi\rangle \langle \psi|
\]

\noindent Tracing out the environment modes:

\[
\rho_{A,A'} = \text{Tr}_{\{{E, F, G, E', F', G'}\}} (\rho)
\]

\noindent where the partial trace is taken over the environment modes.

\noindent The probability of success of Charlie's BSM is given by:

\[
P_C = \operatorname{Tr}(\rho_{A,A'}).
\]

Therefore, the total probability of Charlie's relay is
\[
P_1 = 2P_C,
\]
where the factor of 2 is because there are two successful BSM outcomes, with passive phase correction depending on the outcome. For small $\chi$, we have $P_1 \approx 2 \, \chi^2 \, \eta_C$.

\noindent Finally, the state is renormalised:

\[
\rho_{A,A'} \to \frac{\rho_{A,A'}}{\operatorname{Tr}(\rho_{A,A'})}
\]

In the limit that \( \chi \to 0 \), the state has 1 ebit.

\textbf{Proof:} The dominant terms are:
\begin{itemize}
    \item \( |\psi_1\rangle \) when \( n=0 \) and \( m=1 \),
    \item \( |\psi_2\rangle \) when \( n=1 \) and \( m=0 \).
\end{itemize}

The full state in the limit \( \chi \to 0 \) is given by:
\begin{align*}
\lim_{\chi\to0} |\psi\rangle =     |\psi_1\rangle  \text{ when }  n=0 \text{ and }  m=1 \\ + 
      |\psi_2\rangle  \text{ when }  n=1 \text{ and }  m=0  \\ = \frac{ |0,0,0,0,1,0,0,0\rangle + |1,0,0,0,0,0,0,0\rangle }{\sqrt{2}},
\end{align*}
\noindent keeping only the dominant terms, where the modes are labeled as \(\{ A, E, F, G, A', E', F', G' \}\). We see that the environment modes are all zero and there are no losses. 

That is, we have
\[
\lim_{\chi\to0} |\psi\rangle = |\psi\rangle_{A,A'} = \frac{|1,0\rangle + |0,1\rangle}{\sqrt{2}}
\]
which is a pure maximally entangled Bell state shared between Alice and Bob.

We have analysed the entanglement generation process via sequential Bell-state measurements. For small $\chi$, the first BSM teleports Alice's mode into the channel with success probability $P_0 \approx \eta_1/\sqrt{2}$, and the second BSM, performed on two distributed copies, succeeds with probability $P_1 \approx 2\chi^2 \eta_C$. The resulting renormalised state is a high-fidelity Bell pair,

\[
\ket{\psi}_{A,A'} \approx \frac{\ket{1,0} + \ket{0,1}}{\sqrt{2}}.
\]

\section{Switching loss}\label{sec:switching_loss}

\begin{figure}
\centering \includegraphics[width=0.5\textwidth]{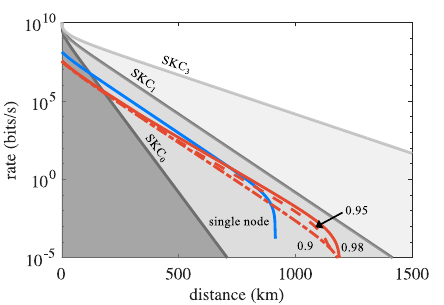} \caption{Key rates vs. total distance comparing single node protocol (blue), capacities (grey), and multi-node protocol (red), for quantum memory loss rate 0.2 dB/km, showing different switching efficiencies (as marked in the figure).}
\label{fig:protocol_switching_efficiencies}
\end{figure}

In \cref{fig:protocol_switching_efficiencies}, we compare switching efficiencies at a fixed memory loss rate $\alpha_{\text{QM}} = 0.2$ dB/km, with all other parameters as in Fig.~2(b) in the main text.

Every time a switch is used it experiences the lossy channel with the transmissivity marked. The rate is optimised over the number of switches used so automatically chooses the best number of times to use a switch.
 
\section{Optimal relay position}\label{sec:relay_position}

\begin{figure}
\centering \includegraphics[width=0.5\textwidth]{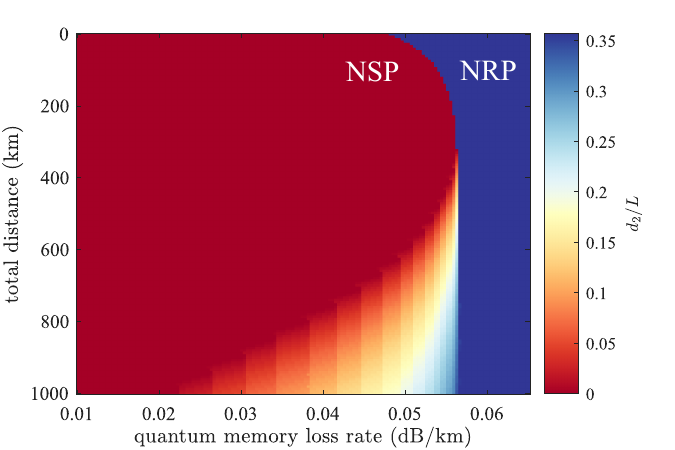} \caption{Heatmap of the optimal $d_2/L$ ratio as a function of quantum memory loss rate $\alpha_{\text{QM}}$ and total distance $L$. This ratio determines optimal node placement for maximizing key rates. Regions correspond to ``node receives photon'' (NRP) and ``node sends photon'' (NSP) configurations. The blue–red boundary marks the break-even point ($\alpha_{\text{QM}} < 0.056$ dB/km) for efficient repeater operation. Ideal, lossless switches are assumed.}
\label{fig:d2}
\end{figure}

\Cref{fig:d2} shows how quantum memory loss affects optimal node placement, with a heatmap of optimal $d_2/L$ maximizing the key rate as a function of $\alpha_{\text{QM}}$ and total distance $L$, assuming ideal switches. 

In the red region, where memory loss is low, repeater-like operation becomes viable—marking the break-even point where repeater gains outweigh losses. Reducing memory loss (e.g., via error correction) is crucial. While ultra-low-loss fibres and fast classical communication can surpass single-node protocols, reaching the three-repeater bound requires memory loss below $\approx$ 0.056 dB/km. This numerically determined threshold for the single-rail protocol is remarkably close to the analytical result obtained for the dual-rail protocol derived above (\( 2\alpha/7 \approx 0.057\) dB/km).

\section{Storage and synchronisation}\label{sec:syncronisation}

Each prepared state is stored in a fixed optical buffer of length \( t_{\text{QM}_1} \), represented as a loop in Fig.~2(a), until classical information about the relay's success or failure is received from the nearby node. If the relay fails, an optical switch directs the pulse to be discarded, as indicated by the ``fail'' label in the diagram. 

If the relay succeeds, the state is stored either in a fixed optical buffer of length \( t_{\text{QM}_2} \) or in a variable optical buffer with a storage time less than \( t_{\text{QM}_2} \). The choice depends on whether a state is already stored in the other fixed buffer. If a state is stored in the other fixed buffer, the variable optical buffer temporarily stores the light for a duration less than \( t_{\text{QM}_2} \), ensuring that the light remains synchronised.

The maximum optical loss due to the fibre lengths, is simply given by the total storage time in the buffers which is \( t_{\text{QM}_1} + t_{\text{QM}_2} \). The worst-case number of switches required is 
\[
1 + \text{max. no. of switches for the variable optical buffer.}
\]

This value can be expressed as:
\[
\text{Max. number of switches} = b \cdot \lceil \log_b(m) \rceil + 1
\]
where \( b \) is the base of the buffer system, and \( m \) is the maximum number of storage steps.

The reasoning behind this formula relates to the number of digits in \( m \). For example, consider a scenario where a state must be stored for a maximum of \( m \) steps. In the worst case, for an \( n \)-digit number in base \( b \), the maximum number of switches corresponds to \( n \cdot b \). To determine \( n \), the number of digits in \( m \), we compute \( \lceil \log_b(m) \rceil \) and multiply it by \( b \). Adding one accounts for the initial fixed optical switch.

Therefore, the total transmissivity in the local optical quantum memories/buffer is given by:
\[
\eta_\text{QM} = \eta_{\text{QM}_1} \eta_{\text{QM}_2} \eta_\text{switch}^{b \lceil \log_b(m) \rceil + 1},
\]
where \( \eta_{\text{QM}_1} \) and \( \eta_{\text{QM}_2} \) represent the transmissivity in the fixed optical buffers, and the final term accounts for the worst-case transmissivity due to the variable optical buffer and switching operations (i.e., using the switch ${b \lceil \log_b(m) \rceil + 1}$ times).

\section{Wait and storage times}\label{sec:wait_times}

\begin{figure}
\centering \includegraphics[width=0.65\textwidth]{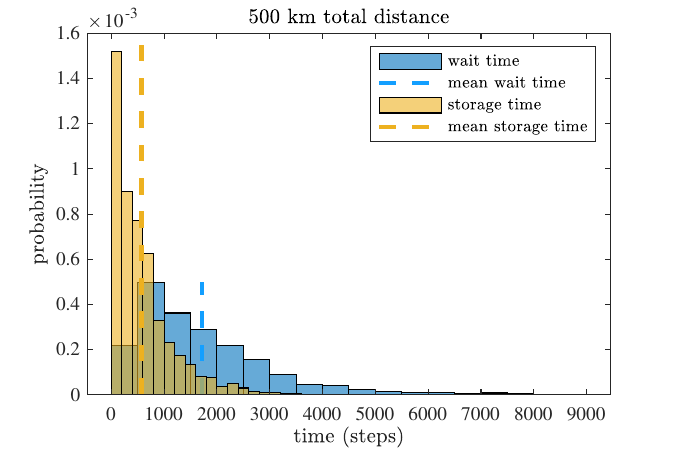}
 \caption{Normalised histograms for the wait and storage times.}\label{fig:storage_times}
\end{figure}

We simulate the waiting and storage times during the protocol. The results are shown in \cref{fig:storage_times}.

\begin{figure}
\centering
\includegraphics[width=0.65\textwidth]{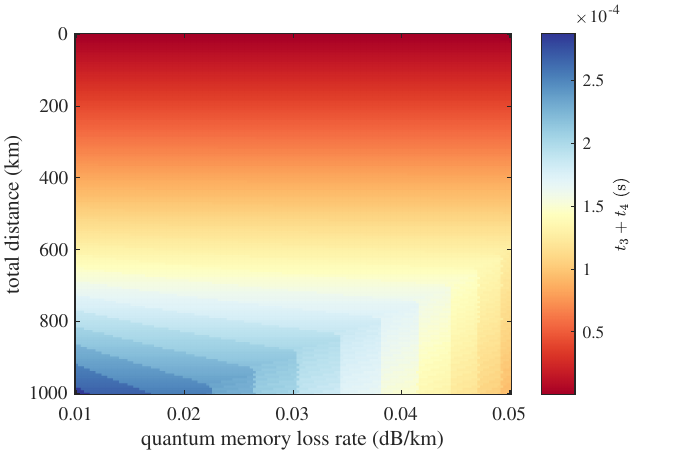}
\caption{Total optical storage time in the quantum repeater configuration. Entanglement is placed at the repeater node ($d_2=0$), and storage times are dominated by the two-way classical communication delay. This configuration achieves scaling approaching \(\eta^{0.25}\), but only with very low loss rates (\(\ll 0.5 \ \mathrm{dB/km}\)) and extended storage times (\(100 \ \mu\mathrm{s}\)).}\label{fig:quantum_repeater_storage}
\end{figure}

\begin{figure}
\centering
\includegraphics[width=0.65\textwidth]{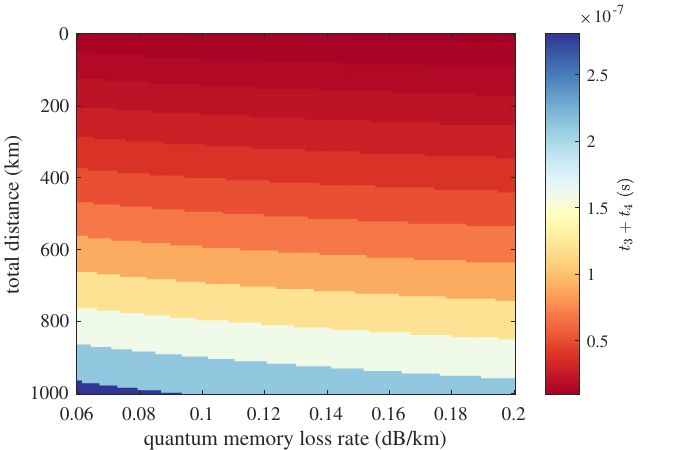}
\caption{Total optical storage time in the slow-light-speed through fibre configuration. Entanglement is placed further into the channel ($d_2>0$), away from the central repeater node. This configuration eliminates classical communication delays, reducing storage times significantly. TF QKD can be surpassed with storage times of \(100 \ \mathrm{ns}\), though achieving \(\eta^{0.25}\) scaling is not possible.}\label{fig:slow_light_storage}
\end{figure}

We examine the total time light needs to be stored in the optical memory for two configurations: the quantum repeater configuration (\cref{fig:quantum_repeater_storage}) and the slow-light configuration (\cref{fig:slow_light_storage}). These plots show the total storage time as a function of the total distance (y-axis) and the loss rate (x-axis), highlighting the trade-offs between wait times, loss rates, and performance.

In the quantum repeater configuration (\cref{fig:quantum_repeater_storage}), entanglement is established at the repeater node. This configuration requires extended storage times due to the necessity of two-way classical communication. For example, achieving a scaling of \(\eta^{0.25}\) demands low loss rates (\(< 0.5 \ \mathrm{dB/km}\)) and storage times of approximately \(100 \ \mu\mathrm{s}\).

In contrast, the slow-light configuration (\cref{fig:slow_light_storage}) leverages the low velocity of light in optical fibres to place the entanglement further into the channel, away from the central repeater node. This eliminates the need for classical communication delays, reducing storage times significantly. For instance, we show that TF QKD can be surpassed with storage times of only \(100 \ \mathrm{ns}\) in this regime, though achieving the \(\eta^{0.25}\) scaling is not possible in this configuration.

These results demonstrate the critical role of storage times and loss rates in determining the viability of different quantum communication strategies, particularly when scaling beyond TF QKD.

\newpage

\noindent [S1] M. S. Winnel, J. J. Guanzon, N. Hosseinidehaj, and T. C. Ralph, Overcoming the repeaterless bound in continuous-variable quantum communication without quantum memories (2021), arXiv:2105.03586 [quant-ph].

\noindent [S2] R. Garc\'ia-Patr\'on, S. Pirandola, S. Lloyd, and J. H. Shapiro, Phys. Rev. Lett. \textbf{102}, 210501 (2009).

\noindent [S3] E. Shchukin, F. Schmidt, and P. van Loock, Phys. Rev. A \textbf{100}, 032322 (2019).




\end{document}